\documentclass[12pt, draftclsnofoot,onecolumn]{IEEEtran}
\usepackage{dsfont}
\usepackage{amssymb}
\usepackage{float}
\usepackage{subfig}
\usepackage{graphicx, amssymb, amsmath}
\usepackage{extarrows}
\usepackage{algorithm} 
\usepackage{algorithmic} 
\usepackage{multirow} 
\usepackage{color}
\usepackage{cite}
\emergencystretch=\maxdimen
\IEEEoverridecommandlockouts
\begin{document}

\title{Air-Ground Integrated Vehicular Network Slicing with Content Pushing and Caching}

\author{Shan~Zhang,~\IEEEmembership{Member,~IEEE,}
	Wei~Quan,~\IEEEmembership{Member,~IEEE,}
	Junling~Li,~\IEEEmembership{Student~Member,~IEEE,}
	Weisen~Shi,~~\IEEEmembership{Student~Member,~IEEE,}
	Peng~Yang,~\IEEEmembership{Student~Member,~IEEE,}
	and~Xuemin~(Sherman)~Shen,~\IEEEmembership{Fellow,~IEEE}
	\thanks{Shan~Zhang is with Beijing Key Laboratory of Computer Networks, the School of Computer Science and Engineering, Beihang University, Beijing, 100191, P.R.China. (Email: zhangshan18@buaa.edu.cn).}
	\thanks{Wei Quan is with the School of Electronic and Information Engineering, Beijing Jiaotong University, Beijing, China (email: weiquan@bjtu.edu.cn).}
	\thanks{Junling~Li, Weisen~Shi and Xuemin~(Sherman)~Shen are with the Department of Electrical and Computer Engineering, University of Waterloo, 200 University Avenue West, Waterloo, Ontario, Canada, N2L 3G1 (email: \{j742li, w46shi, sshen\}@uwaterloo.ca).}
	\thanks{Peng Yang is with the School of Electronic Information and Communications, Huazhong University of Science and Technology, Wuhan, China (email: yangpeng@hust.edu.cn).}
}

\maketitle

\begin{abstract}

	In this paper, an Air-Ground Integrated VEhicular Network (AGIVEN) architecture is proposed, where the aerial High Altitude Platforms (HAPs) proactively push contents to vehicles through large-area broadcast while the ground roadside units (RSUs) provide high-rate unicast services on demand.
	To efficiently manage the multi-dimensional heterogeneous resources, a service-oriented network slicing approach is introduced, where the AGIVEN is virtually divided into multiple slices and each slice supports a specific application with guaranteed quality of service (QoS).
	Specifically, the fundamental problem of multi-resource provisioning in AGIVEN slicing is investigated, by taking into account typical vehicular applications of location-based map and popularity-based content services.
	For the location-based map service, the capability of HAP-vehicle proactive pushing is derived with respect to the HAP broadcast rate and vehicle cache size, wherein a saddle point exists indicating the optimal communication-cache resource trading.
	{{For the popular contents of common interests, the average on-board content hit ratio is obtained, with HAPs pushing newly generated contents to keep on-board cache fresh.
	Then, the minimal RSU transmission rate is derived to meet the average delay requirements of each slice.}}
	The obtained analytical results reveal the service-dependent resource provisioning and trading relationships among RSU transmission rate, HAP broadcast rate, and vehicle cache size, which provides guidelines for multi-resource network slicing in practice.
	Simulation results demonstrate that the proposed AGIVEN network slicing approach matches the multi-resources across slices, whereby the RSU transmission rate can be saved by 40\% while maintaining the same QoS.
	
\end{abstract}


\section{Introduction}

	Air-ground integrated networks are expected to provide high capacity with seamless coverage by exploiting the complementary advantages of dense terrestrial infrastructure and large-coverage aerial stations \cite{Grace05_HAP_broadcast_integrated}. 
	Specifically, the High Altitude Platforms (HAPs), usually airships or balloons operating in the stratosphere {{at an altitude of}} 17-22 km, can provide broadcast services to effectively relieve the network load on ground \cite{Djuknic97_HAP_early}.
	Since the 1990s, HAPs have been proposed to support the digital video/audio broadcast services, which has been proved to be feasible and energy efficient compared with the terrestrial systems \cite{Mohammed11_HAP_role_proceeding}.
	As data streaming now becomes the dominant mobile service, HAP broadcast technologies can facilitate more applications by exploiting the concentrated requests of popular contents \cite{5G_Overview_JSAC14_JAndrews}.
	
	{{Equipped with on-board communication modules, vehicles are envisioned as the future fourth screen raising enriched mobile applications for navigation, entertainment, and road safety enhancement \cite{Chen18_VANET_capacity_TVT,Abboud16_DSRC_cellular_interwork}.}}
	In this regard, additional network infrastructures are expected to be further deployed, since the existing ground networks are insufficient to accommodate these emerging data-hungry and delay-sensitive applications \cite{weisen18_drone_netw,Ning17_SAG_net,Mao16_cooperative_offloading_TVT}.
	As vehicular mobile applications are typically location-based and more predictable, the large-coverage HAP broadcast is a promising solution to relieve the ground mobile traffic, by exploiting the storage for on-board caching \cite{mine_VeCache_COMMAG}.
	
	In this paper, we propose an air-ground integrated vehicular network (AGIVEN) architecture, whereby the HAPs broadcast popular contents to vehicles \emph{a priori} to requests while the ground roadside units (RSUs) provide services on demand through unicast.
	With contents cached on board, vehicles can enjoy low latency and enhanced RSU access rate with the reduced traffic load.
	{{Despite the attractive advantages, the AGIVENs demonstrate multi-dimensional heterogeneity of both resources and traffic demands, posing significant challenges to network management.
	To address this issue, the service-oriented network slicing method is applied.
	Based on the network virtualization concept, network slices are constructed on the top of physical resources, where each slice supports a specific service with the guaranteed quality of service (QoS).}}
	In this work, three slices are constructed for the typical vehicular applications, i.e., high definition (HD) map for navigation (MaNa) slice, file of common interest (FoCI) slice, and on-demand transmission (ODT, such as interactive sessions, voice call and web viewing) slice.
	The MaNa slice has more strict delay requirement compared with that of the FoCI slice, and both slices are push-enabled considering the concentrated and predictable requests \cite{Wang16_cache_mobility_mag}.
	The ODT slice is rather elastic, which is served only by RSUs with the remaining transmission resources. 
	
	Two fundamental problems of the AGIVENs are explored, (1) how to slice the multi-resources to accommodate the three applications efficiently, and (2) how much on-ground traffic can be self-served with HAP proactive content pushing.	
	The main challenges exist in the service-dependent heterogeneous resource trading.
	To address these issues, we analyze the service capability of MaNa and FoCI slices with respect to HAP broadcast rate, vehicle cache size and RSU transmission rate, taking into account vehicle mobility and content popularity features. 
	As the MaNa slice requires location-based services whereas the FoCI slice are served according to content popularity distribution, the two slices are analyzed separately in different ways.
	
	{{In the MaNa slice, vehicles cache maps of the successive blocks on the route through the HAP broadcast, and will turn to the RSUs to download the remaining map segments of the current block out of the cache.}}
	To study the offloading capability of HAPs, we derive the probability that a vehicle has cached the complete map before entering the corresponding block, defined as the \textit{accomplishment ratio}.
	The accomplishment ratio is proved to be an increasing function of HAP broadcast rate, where the increasing rate first increases and then decreases.
	The saddle point of HAP broadcast rate shows an inversely linear form with respect to the vehicle cache size, suggesting the optimal match of communication and storage resources in the MaNa slice.
	{{In addition, the RSU service (i.e., remaining map downloading) process is analyzed based on queueing models, whereby the delay-constrained RSU transmission rate requirement is derived with a conservative approximation.}}
	Analytical results show that the required RSU rate is a convex decreasing function of the accomplishment ratio, which demonstrates the three-dimensional (3D) resource trading relationship among HAPs, RSUs, and vehicles.
	
	In the FoCI slice, vehicles cache the most popular contents on board for potential future use.
	As contents can generate and expire randomly, HAPs broadcast the newly generated files to vehicles to maintain the on-board content hit ratio.
	The number of valid files cached on-board is time-varying, which is modeled as a Markovian birth-and-death process to analyze the steady-state probability.
	Then, the average content hit ratio is derived, depending on the normalized file update rate, i.e., the ratio of HAP-assisted vehicle cache update rate to the file expiration rate. 
	Furthermore, the delay-constrained RSU transmission rate of FoCI slice is also derived with respect to HAP transmission rate and vehicle cache size, revealing the 3D resource trading relationship in the FoCI slice.
	In specific, the performance of the FoCI can fall into two regions: (1) communication-constrained region, where the low HAP broadcast rate degrades the vehicle cache utilization, or (2) cache-constrained region, where the HAP broadcast contents cannot be stored due to cache overflow.
	
	{{The analytical results of MaNa and FoCI slices both reveal the necessity to optimize slice-level resource matching in network slicing.}}
	Specifically, the resource matching relationships are service-dependent.
	For instance, for the given HAP broadcast rate and vehicle cache size, the optimal RSU transmission rate should increase with vehicle mobility in the MaNa slice, whereas it varies with the content popularity distribution in the FoCI slice.
	Extensive simulations are conducted to validate the theoretical analysis on slice-level resource provisioning, trading and matching.
	In addition, numerical results demonstrate that the resources of RSUs required by the MaNa and FoCI slices can be saved up to 40\% with HAP-vehicle proactive content pushing and caching, through inter-slice resource matching.
	
	The main contributions of this paper are as follows:
	
	1) An air-ground integrated vehicular network framework is proposed, where the HAPs exploit the content information and offload ground traffic through broadcast;
	
	2) Network slicing problem is investigated with the multi-dimensional heterogeneous resources (i.e., HAP broadcast rate, RSU unicast rate, and vehicle cache size) and differentiated services (location-based MaNa, and popularity-based FoCI);
	
	3) The offloading capability of HAP pushing is derived in closed form for the MaNa and FoCI slices, respectively; and the delay-constrained RSU, HAP, and vehicle resource trading relationships are obtained in an analytical way; 
	
	4) The obtained results reveal the optimal resource matching and sharing among the MaNa, FoCI, and ODT slices, which can be applied for cost-effective AGIVEN deployment and management in practical systems. 

	
	
	The remaining of this paper is organized as follows. 
	Section~\ref{sec_review} reviews the state-of-the-art research on HAPs, pushing and caching, and network slicing. Section~\ref{sec_system_model} introduces the system model of AGIVENs, based on which the MaNa and FoCI slices are analyzed in Sections~\ref{sec_map_analysis} and \ref{sec_file_analysis}, respectively. Section~\ref{sec_simulation} provides simulation results, and Section~\ref{sec_conclusions} draws conclusions.

\section{Literature Review}
	\label{sec_review}
	Located quasi-statically in the stratosphere, the HAP-based stations enjoy large coverage area, high-probability Line of Sight (LoS) links, flexible and dynamic deployment.
	Thus, constructing HAP-based airborne networks has been considered as a promising solution to enhance the ground communication systems in both coverage and capacity.
	{{To exploit these advantages, extensive studies have been conducted from aspects of the Air-to-Ground (A2G) channel modeling \cite{Hourani14_A2G_pathloss_letter}, aerial network architecture design \cite{Chandrasekharan16_HAP_LTE_A_mag, Dong16_HAP_constellation_TMM, Yaliniz16_multi_tier_HAP, Alzenad17_all_HAP}, aerial network deployment \cite{LiuLiang18_UAV_placement}, and resource management \cite{Foo00_HAP_interference_cancellation,Grace05_HAP_spectrum_sharing,Foo02_HAP_power_management}.}}
	Chandrasekharan et al. have proposed to implement LTE-A in aerial platforms to provision Internet access during temporary events and emergencies \cite{Chandrasekharan16_HAP_LTE_A_mag}.
	Dong et al. have further investigated the cost-efficient HAP constellation deployment, aiming at maximizing network capacity under the QoS constraints \cite{Dong16_HAP_constellation_TMM}.
	In addition, a multi-tier heterogeneous aerial network consisting of HAPs and Low Altitude Platforms (LAPs) (like drones) has been designed in \cite{Yaliniz16_multi_tier_HAP}, where the HAPs form the core of airborne and conduct functions like congestion control.
	As an extension of \cite{Yaliniz16_multi_tier_HAP}, Alzenad et al. have leveraged all types of flying platforms including HAP, Middle Altitude Platform (MAP), and LAPs to provide communication services through free-space optics (FSO) based A2G links \cite{Alzenad17_all_HAP}.  
	HAP resource management schemes have been designed for interference cancellation \cite{Foo00_HAP_interference_cancellation}, spectrum sharing \cite{Grace05_HAP_spectrum_sharing}, and power allocation \cite{Foo02_HAP_power_management}.
	However, these studies focused on the HAP unicast, while HAP broadcast has been mainly applied to Digital Video/Audio Broadcast (DVB/DAB) \cite{karapantazis05_HAP_survey}.
	Different from the existing works, this paper proposes to apply HAP broadcast-based proactive pushing and vehicle caching to support differentiated vehicular mobile applications, through multi-dimensional heterogeneous network slicing and sharing.
	
	Mobile edge caching exploits the storage resources of base stations or access points to relieve the backhaul transmission, which can reduce end-to-end delay and overcome backhaul constraints \cite{Bastug14_cache_framework_BS_D2D_mag}.
	{{Furthermore, proactive content pushing exploits the storage resource of end users to reduce duplicated wireless transmission, which provides an effective way to improve network capacity, energy efficiency, and QoS \cite{Gong17_push_TWC,Li11_push_JSAC,Si16_push_video_JSAC}.}}
	In fact, mobile edge caching and proactive content pushing both trade communication resources with storage resources, and the theoretical relationship has been analyzed for the ground cellular networks \cite{mine_cache_TMC, mine_caching_TVT, Feng16_push_update_TWC}.
	Although insightful, the existing works still focus on single service, and do not consider to balance communication and storage resource among different traffic demands.
	To address this issue, the network slicing method can be implemented \cite{junling_survey,Junling17_slicing_GC}. 
	The network slicing approaches have been studied for ground radio access networks (RANs), which can be classified into four levels (i.e., spectrum-level, infrastructure-level, network-level, and flow-level) \cite{junling_survey}. 
	A cellular base station is sliced with resource reservation to maximize the total revenue while satisfying the specific requirements of each slice \cite{junling_NVS}. 
	Network slicing also enables RAN sharing among multiple operators \cite{junling_cellslice, junling_multitenant}, wherein flow control and resource allocation schemes have been designed to guarantee fairness with QoS isolation. 
	The very recent work \cite{junling_information-centric} has considered content caching in network slicing design, where 
	the virtual resource allocation and in-network caching have been jointly designed by solving an optimization problem, aiming at maximizing the aggregate network utility.
	{{Compared with the existing works, this paper goes one step further on network slicing, considering the multi-dimensional heterogeneous resource and differentiated services requirements.
			In addition, the advanced proactive HAP pushing and vehicle caching are employed to effectively reduce duplicated on-ground transmissions. }}

\section{System Model}
    \label{sec_system_model}
    
    
    
\begin{figure}[!t]
    	\centering
    	\includegraphics[width=3.5in]{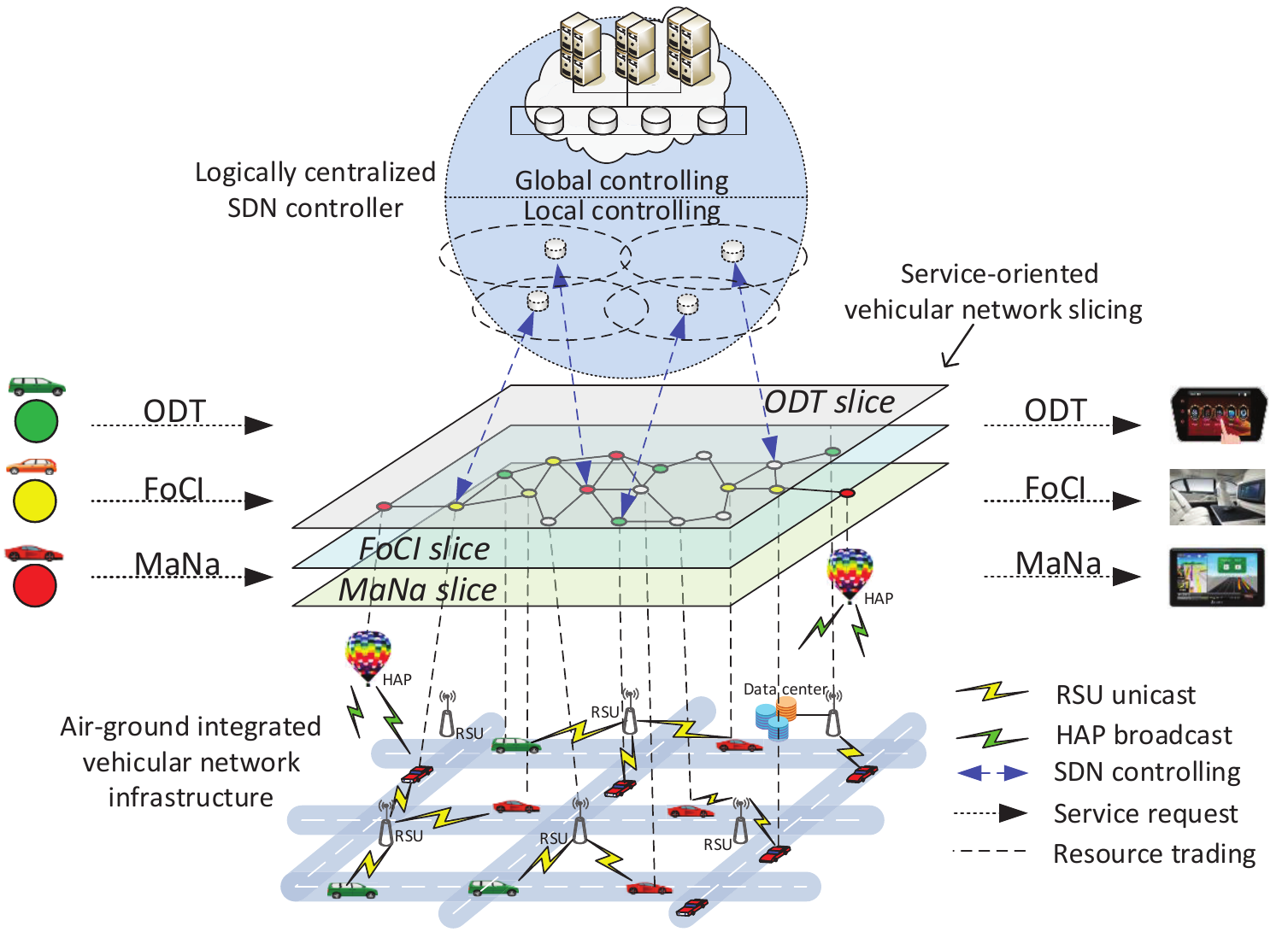}
    	\caption{Service-oriented network slicing in AGIVENs.}
    	\label{fig_slicing}
    \end{figure}

    A typical AGIVEN scenario is illustrated in Fig.~\ref{fig_slicing}, which consists of sparse aerial HAPs and dense ground RSUs. 
    Each HAP can form multiple cells through advanced beamforming technologies, for efficient spatial spectrum reuse \cite{Thornton03_HAP_cell, Dessouky07_HAP_cell}. 
    {{The HAPs are considered to have ideal uplink with ground stations and can update contents timely, yet the downlink broadcast rate is constrained due to the intensive traffic demand and limited bandwidth.}}
    RSUs are uniformly distributed on the ground, where each covers a block and provides unicast service upon vehicle user requests.
    Three typical on-road mobile applications are supported, i.e., the delay-sensitive MaNa, FoCI of medium delay requirement, and the ODT which prefers higher data rates.
    
    {{To support the differentiated services, a software defined network (SDN) based hierarchical controller is built to construct and manage virtual slices on-the-top of network resources \cite{QuanWei_SDN_vehicle}.
    		Local controllers are placed at all HAPs and RSUs to conduct resource virtualization, whereby {{the physical resources (e.g., channels and power) are abstracted as logical resources (e.g., transmission rate) based on the measured channel conditions}}.
    		Meanwhile, the local controllers monitor mobile traffic demand, and send reports to the central controller.
    		The main role of central controller is two-fold, i.e., to determine the logical resources slicing through the southbound interface, and to meet the QoS requirement of each slice through the northbound interface.
    		The slicing decisions are transmitted to local controllers for physical resource and user scheduling.
    		Then, the central controller can adjust the slicing decision based on the QoS performance of each slice.}}
    In this work, we design the multi-resource slicing algorithm at the central controller, considering the downlink performance within one HAP cell coverage.
    Specifically, the HAP broadcast rate, RSU unicast rate, and vehicle cache size are jointly sliced, aiming at maximizing the rate of ODT slice while satisfying the delay requirements of MaNa and FoCI slices.

    \subsection{MaNa Slice Modeling}
    
\begin{figure}[!t]
    	\centering
    	\includegraphics[width=3in]{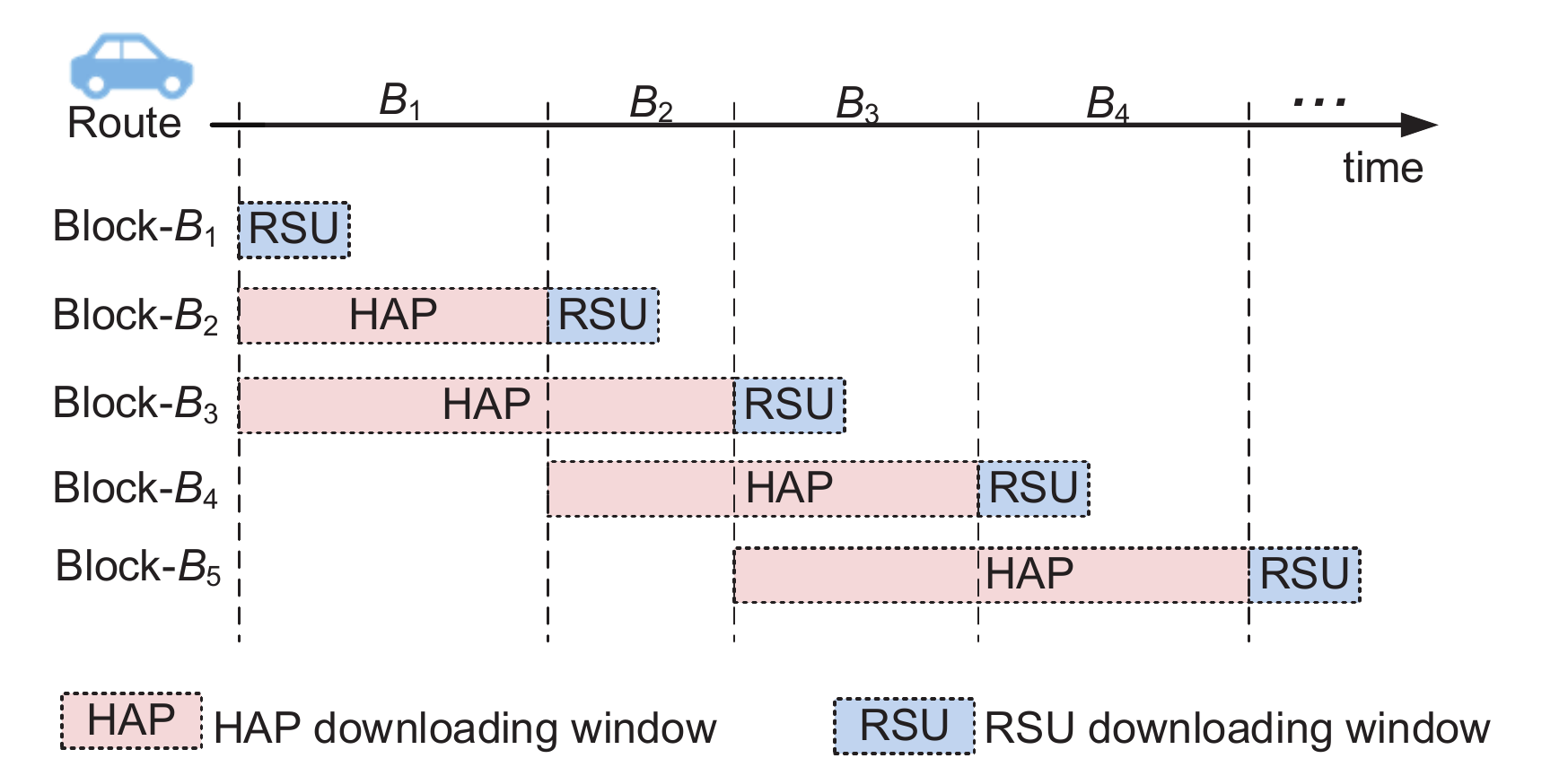}
    	\caption{An illustration of MaNa service process: each vehicle can cache map files of two blocks.}
    	\label{fig_map_process}
    \end{figure}

    {{The Manhattan mobility model is adopted for vehicle mobility \cite{Manhattan_model_2},}} where each road block generates a map file of the same size $L_\mathrm{m}$.
    The HAP broadcasts the maps of all blocks within coverage through radio resource sharing, while each vehicle only downloads the maps on demand.
    {{Consider a vehicle driving on route $[B_1,B_2,...B_J,...]$, where $B_J$ denotes the $J$th block. }}
    Denote by $C_\mathrm{m}$ the normalized vehicle cache size for the map, i.e., the maximal number of maps can be cached on board.
    The map downloading process is illustrated in Fig.~\ref{fig_map_process}.
    Due to the constrained cache size, the vehicle only maintains the maps of $C_\mathrm{m}$ proceeding blocks on route to avoid cache overflow. 
    {{Therefore, the vehicle starts to cache the map of $B_J$ when it enters $B_{J-C_\mathrm{m}}$, and stops downloading under either of the two conditions: (1) the map of $B_J$ has been downloaded into cache completely, and (2) the vehicle enters $B_J$.
    		Notice that there exists a valid HAP downloading window for each block, as shown in Fig.~\ref{fig_map_process}.
    		Accordingly, the vehicle downloads the maps of $[B_{J+1},B_{J+2},...,B_{J+C_\mathrm{m}}]$ from the HAP when driving on $B_{J}$.}}	
    
    According to Fig.~\ref{fig_map_process}, the available HAP window size to download the map of $B_J$ is 
    \begin{equation}
    	\label{eq_window}
    	T_J^{+} = \sum_{j=\max\{1,J-C_\mathrm{m}\}}^{J-1} t_{j},
    \end{equation}
    where $t_j$ is the dwelling time on $B_J$.
    Then,	
    the remaining map downloaded from the RSU on $B_J$ is given by
    \begin{equation}
    	\label{eq_remain_map_size}
    	L_J^{-} = \max\{0, L_\mathrm{m} - R_\mathrm{HM} T_J^{+}\},
    \end{equation}	
    where $R_\mathrm{HM}$ is HAP broadcast rate.
    Note that $L_J^{-}$ is a random variable, since the window size $T_J^{+}$ has two-dimensional randomness of block number $J$ and dwelling time $t_j$. 
    
    Suppose vehicles enter the road following Poisson process of parameter $\lambda_\mathrm{v}$, reflecting the inter-vehicle headway and speed \cite{Mirchandani07_Poisson_vehicle}.
    The dwelling time on one block can be modeled as Erlang distribution of parameter $K$ and $\mu$, i.e., $\mathrm{E} (K, \mu_\mathrm{v})$.
    Accordingly, the mean and variance of block-level dwelling time are $K/\mu_\mathrm{v}$ and $K/{\mu_\mathrm{v}}^2$, respectively \cite{Kobayashi99_vehicle_Erlang_IEICE, Fang99_cellular_mobility_erlang}.
    By adjusting parameters $\lambda_\mathrm{v}$, $K$ and $\mu$, the Erlang model of dwelling time can simulate specific scenarios such as rural rush hours (by increasing $\lambda_v$ and $K$, or reducing $\mu$).
    {{As vehicles enter a block following Poisson process and request to download the remaining map files out of cache, the service process of the corresponding RSU can be modeled as an $M/G/1$ queue.}}
    The service rate is $R_\mathrm{RM}/L_J^{-}$, where $R_\mathrm{RM}$ is the RSU transmission rate for the MaNa slice.
    The average delay of map downloading (consisting of both queueing and transmission) should be guaranteed no larger than a threshold for driving safety and efficiency.
    To this end, sufficient resource should be provisioned, and detailed analysis will be presented in Section~\ref{sec_map_analysis}. 

    \subsection{FoCI Slice Service and Modeling}
    
    \begin{figure}[!t]
    	\centering
    	\includegraphics[width=2in]{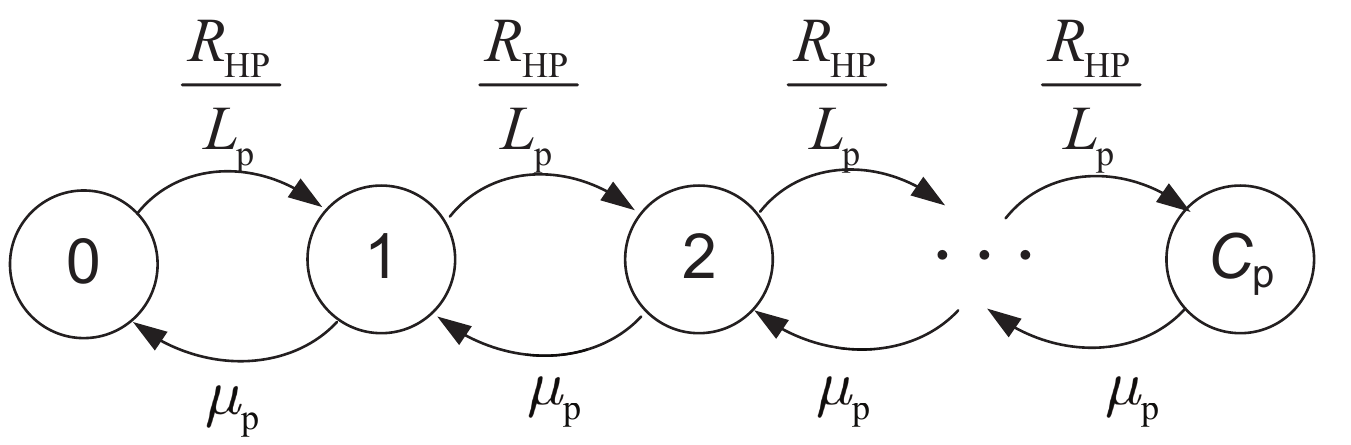}
    	\caption{The state transition of FoCI slice, i.e., the number of valid files in vehicle cache.}
    	\label{fig_FoCI_status}
    \end{figure}
    
    In the FoCI slice, vehicles cache the most popular contents.
    When a vehicle requests a content, it will be self-served if the content is on board, and otherwise, send a request to the associated RSU for service.
    Define the on-board content hit ratio as the probability that the vehicle can be self-served, which is a key performance metric of the FoCI slice.	
    In practical systems, the cached contents may expire as new contents generate.
    HAP-assisted cache update is conducted, where the HAP broadcasts newly generated popular contents at rate $R_\mathrm{HP}$.
    Suppose the life time of a popular file follows the exponential distribution of mean $1/\mu_\mathrm{p}$ \cite{Feng16_push_update_TWC}.
    Denote by $L_\mathrm{p}$ the file length, and $C_\mathrm{p}$ the cache size normalized by file length.	
    The number of valid files in vehicle cache varies with random file expiration and update.
    The process is modeled as a Markovian chain as shown in Fig.~\ref{fig_FoCI_status}.
    The state denotes the number of valid files in cache, and the time to update a file is considered to follow exponential distribution \cite{Feng16_push_update_TWC}.
    The average hit ratio can be obtained based on the steady-state probability analysis of Markovian chain.
    
    Assume that FoCI requests arrive following Poisson process of rate $\lambda_\mathrm{p}$, and are served in a First-in-First-Out (FIFO) manner with a constant rate $R_\mathrm{RP}$.
    The service process of FoCI at the RSU can be modeled as {{an M/D/1 queue}} with arrival rate of $\lambda_\mathrm{p} (1-P_\mathrm{hit})$, where $P_\mathrm{hit}$ is the on-board hit ratio.
    Accordingly, the on-board hit ratio reflects the offloading capability of HAP-vehicle caching, and the delay performance of FoCI will be analyzed in Section~\ref{sec_file_analysis}.

    
    \subsection{Resource Slicing and Sharing Modeling}
    
    As the utility of ODT slice increases with the transmission rate, the AGIVEN slicing problem can be formulated as:
    
    \begin{subequations}
    	\small
    	\label{eq_P1}
    	\begin{align}
    		\max\limits_{\tiny{\left\{\begin{array}{l} C_\mathrm{m}, R_\mathrm{HM},R_\mathrm{RM}, \\ C_\mathrm{P},R_\mathrm{HP},R_\mathrm{RP}\end{array}\right\}}}~~~& R_\mathrm{R} - R_\mathrm{RM} - R_\mathrm{RP} \\
    		\mbox{s.t.}~~~~~~~& \bar{W}_\mathrm{M} \leq \hat{T}_\mathrm{m}, \\
    		(\mbox{P1})~~~~~~~~~~~~~~~~~~~~~~~~& \bar{W}_\mathrm{P}  \leq \hat{T}_\mathrm{p}, \\
    		& L_\mathrm{m} C_\mathrm{m} + L_\mathrm{p} C_\mathrm{p} \leq L_\mathrm{v},\\
    		& N_\mathrm{block} R_\mathrm{HM} + R_\mathrm{HP} \leq R_\mathrm{H} ,
    	\end{align}
    \end{subequations}	
    {{where $R_\mathrm{R}$ and $R_\mathrm{H}$ are the abstracted transmission rates of an RSU and HAP cell, respectively, $\bar{W}_\mathrm{M}$ and $\bar{W}_\mathrm{P}$ are the average delays of MaNa and FoCI slices, respectively, $\hat{T}_\mathrm{m}$ and $\hat{T}_\mathrm{p}$ are the required average delays of the MaNa and FoCI slices, respectively, $L_\mathrm{v}$ is the vehicle cache size in bit, and $N_\mathrm{block}$ denotes the number of blocks covered by the HAP cell.
    		The objective function of (P1) is to maximize the RSU transmission rate available to the ODT slice; the constraints (\ref{eq_P1}b) and (\ref{eq_P1}c) guarantee the average delay performance of the MaNa and FoCI slices, respectively;}} the constraints (\ref{eq_P1}d) and (\ref{eq_P1}e) are due to the constrained vehicle cache size and HAP transmission rate, respectively. 
    Note that the average delay of the MaNa slice $\bar{W}_\mathrm{M}$ depends on the allocated multi-resources of $C_\mathrm{m}$, $R_\mathrm{HM}$, and $R_\mathrm{RM}$, while the average delay of the FoCI slice $\bar{W}_\mathrm{P}$ depends on $C_\mathrm{p}$, $R_\mathrm{HP}$, and $R_\mathrm{RP}$.
    The challenges of problem (P1) are two-fold: (1) slice-level delay-constrained multi-resource provisioning, i.e., constraints (\ref{eq_P1}b) and (\ref{eq_P1}c); and (2) cross-slice multi-resource sharing and balancing to maximize the objective function.
    We focus on the first issue and conduct in-depth analysis, as presented in Sections~\ref{sec_map_analysis} and \ref{sec_file_analysis}.
    Based on the analytical results, the second problem can be addressed, and we show numerical results as illustrations.
    
\section{MaNa Slice Analysis and Provisioning}
    \label{sec_map_analysis}
	
	This section investigates the performance of MaNa slice with respect to multi-resource provisioning.
	Specifically, we analyze the accomplishment ratio, i.e., the probability that a vehicle can complete map downloading within the HAP downloading window.
	In addition, the average delay at RSUs is obtained by applying the M/G/1 model analysis.
	Furthermore, the upper bounds of RSU delay are also provided through M/D/1 model approximation, revealing the 3D resource trading among HAP broadcast, RSU unicast, and vehicle cache size.
	
	\subsection{Accomplishment Ratio of HAP-Vehicle Map Pushing}
	
	{{The accomplishment ratio of block $B_J$ is given by}}
	\begin{equation}
		P_{\mathrm{acc},J} = \mathds{P} \left\{ T_J^{+} R_\mathrm{HM} \geq L_\mathrm{m} | J \right\},
	\end{equation}
	where $T_J^{+}$ is the HAP window size of $B_J$ given by Eq.~(\ref{eq_window}).
	As $t_j$ follows Erlang distribution of parameter $(K, \mu_\mathrm{v})$, $T_J^{+}$ also follows Erlang distribution of parameter $(K\min\{C_\mathrm{m},J-1\}, \mu_\mathrm{v})$ according to Eq.~(\ref{eq_window}).
	Thus, $T_1^{+} = 0$, and the probability distribution function of $T_J^{+}$ for $J\geq 2$ is given by
	\begin{equation}
		\label{eq_T_pdf}
		P_{T_J^{+}} (t)  = \frac{{\mu_\mathrm{v}}^{K J'}t^{K J'-1} e^{-\mu_\mathrm{v} t}}{(K (J'-1)-1)!},
	\end{equation}	
	where $J'=\min\{J-1, C_\mathrm{m}\}$.
	The accomplishment ratio of $B_J$ is the tail distribution of $T_J^{+}$:
	\begin{equation}
		\label{eq_P_acc_J}
		P_{\mathrm{acc},J} =  1- \frac{\gamma(KJ', x)}{(K J' -1)!},
	\end{equation}
	where $x \triangleq \frac{\mu_\mathrm{v} L_\mathrm{m}}{R_\mathrm{HM}}$, and $\gamma(\cdot,\cdot)$ is the lower incomplete gamma function defined as
	\begin{equation}
		\label{eq_gamma_function}
		\gamma(s,x) = \int_{0}^{x} t^{s-1} e^{-t} \mathrm{d} t.
	\end{equation}
	The physical meaning of $x/K$ is the average normalized HAP window size (measured in blocks) needed to download the complete map in the cache.
	As $\gamma(s, x)$ increases with $x$, the accomplishment ratio increases with the HAP rate $R_\mathrm{HM}$, and decreases with vehicle speed $\mu_\mathrm{v}$.
	
	As vehicle moves, the long-term average accomplishment ratio is given by
	\begin{equation}
		\small
		\label{eq_P_acc}
		\begin{split}
			P_\mathrm{acc} = & \sum_{J=1}^{\infty} G_J P_{\mathrm{acc},J} = \sum_{J=2}^{C_\mathrm{m}} G_J \left( 1-\frac{\gamma(K(J-1),x)}{(K(J-1)-1)!} \right) + \\
			& \sum_{J=C_\mathrm{m}+1}^{\infty} G_J \left( 1-\frac{\gamma(KC_\mathrm{m},x)}{(KC_\mathrm{m}-1)!} \right),
		\end{split}
	\end{equation} 
	where $G_J$ is the probability that a vehicle is passing the $J$-th block.
	$G_J$ depends on the macro-scope vehicle mobility, and we do not constrain the form $G_J$ for general analysis.
	For instance, $G_J=(1-\psi)\psi^{J-1}$ models the case that the route length follows geometric distribution, where a vehicle parks on a block with probability $1-\psi$ and continues driving with probability $\psi$.
	
	{{The accomplishment ratio reflects the offloading capability of HAP-vehicle pushing, and the influence of $R_\mathrm{HM}$ and $C_\mathrm{m}$ should be analyzed to achieve efficient resource provisioning.}}
	However, the analysis is intractable due to the complex addictive form of Eq.~(\ref{eq_P_acc}).
	To reveal insights, the accomplishment ratio is analyzed approximately, based on the upper and lower bounds:
	\begin{subequations}
		\small
		\label{eq_P_acc_bound}
		\begin{align}
			\check{P}_\mathrm{acc} & = \left(\sum_{J=C_\mathrm{m}+1}^{\infty} G_J \right)\left(1-\frac{\gamma(KC_\mathrm{m},x)}{(KC_\mathrm{m}-1)!}\right),\\
			\hat{P}_\mathrm{acc} & = 1-\frac{\gamma(KC_\mathrm{m},x)}{(KC_\mathrm{m}-1)!},
		\end{align}
	\end{subequations}
	where Eq.~(\ref{eq_P_acc_bound}a) comes from the second term of Eq.~(\ref{eq_P_acc}), and Eq.~(\ref{eq_P_acc_bound}b) holds since $\frac{\gamma(s,x)}{(s-1)!}$ decreases with $s$.
	The differences between $\check{P}_\mathrm{acc}$ and $\hat{P}_\mathrm{acc}$ is $\left(\sum_{J=1}^{C_\mathrm{m}} G_J\right) \left(1-\frac{\gamma(KC_\mathrm{m},x)}{(KC_\mathrm{m}-1)!}\right)$.
	Therefore, the upper and lower bounds can be quite tight if the route length is relatively large compared with cache size $C_\mathrm{m}$, i.e., $\sum_{J=1}^{C_\mathrm{m}} G_J \rightarrow 0$.
	Based on (\ref{eq_P_acc_bound}), the relationships between accomplishment ratio, $R_\mathrm{HM}$ and $C_\mathrm{m}$ is demonstrated as Proposition~1.
	
	\textbf{Propositon~1.} The accomplishment ratio of HAP map pushing increases with the HAP broadcast rate. The increasing rate\footnote{{{The increasing rate is the derivative of the accomplishment ratio with respect to the HAP broadcast rate.}}} firstly increases but then decreases, and the saddle point is given by $R^*_\mathrm{HM} =  \frac{1}{KC_\mathrm{m}+2} L_\mathrm{m}\mu_\mathrm{v}$ if $\sum_{J=1}^{C_\mathrm{m}} G_J \rightarrow 0$.
	
	\emph{Proof.}~Please refer to Appendix~\ref{proof_proposition_1}.
	\hfill \rule{4pt}{8pt}\\
	
	According to Proposition~1, the accomplishment ratio increases super-linearly with the HAP broadcast rate if $R_\mathrm{HM} < R^*_\mathrm{HM}$, and increases sub-linearly if $R_\mathrm{HM} > R^*_\mathrm{HM}$.
	This result indicates that the HAP broadcast should be kept in a region around $R^*_\mathrm{HM}$ considering resource efficiency.
	Note that $R_\mathrm{HM} = R^*_\mathrm{HM}$ is equivalent to $\frac{\mu_\mathrm{v} L_\mathrm{m}}{K R_\mathrm{HM}} = C_\mathrm{m} + \frac{2}{K}$.
	The left side is the expected number of blocks passed to complete map downloading from the HAP, while $C_\mathrm{m}$ is the valid HAP window size measured in blocks, shown as Fig.~\ref{fig_map_process}.
	The saddle point indicates that the cache size should match with the downloading speed.
	When $R_\mathrm{HM} < R^*_\mathrm{HM}$, the vehicle cannot fill the cache within the downloading window, degrading the utilization of vehicle cache.
	In this case, increasing transmission rate can improve the accomplishment ratio significantly.
	When $R_\mathrm{HM} > R^*_\mathrm{HM}$, the vehicle can complete map downloading quickly, while further increasing $R_\mathrm{HM}$ no longer improves accomplishment ratio.
	
	\textbf{Remark~1.}~The HAP broadcast rate should be set around saddle point $R^*_\mathrm{HM}$, which is an inversely linear function of the vehicle cache size. 
	
	The important insight is that the cache size can trade communication resources in the linear rational manner.
	In practical systems, $KC_\mathrm{m}$ can be much larger than 2.
	Therefore, the HAP broadcast rate can be considered to decrease linearly with the inverse of cache size $C_\mathrm{m}$.
	
	Furthermore, the vehicle mobility also influences the accomplishment rate.
	As $P_\mathrm{acc}$ increases with $K$ and decreases with $\mu_\mathrm{v}$, the accomplishment ratio decreases with the vehicle speed $K/\lambda$.
	Increasing $K$ or decreasing $\lambda_\mathrm{v}$ is equivalent to increasing the window size of HAP map downloading, and thus improves the accomplishment ratio.
	Specifically, the average duration of window size can be approximated as $KC_\mathrm{m}/\lambda_\mathrm{v}$.
	As the transmission rate can be also traded with cache size, low vehicle mobility can help to reduce the resource consumption of MaNa slice.
	
	\textbf{Remark~2.} The HAP broadcast rate or vehicle cache size provisioned for the MaNa slice can be reduced during traffic jams or in the busy downtown regions.	
	
	\subsection{RSU Service Delay Analysis of MaNa slice}
	
	The mean and variance service time are needed when applying the M/G/1 model to analyze the average waiting time.
	The average service time of the RSU to serve one vehicle with the remaining map is given by
	{{\begin{equation}
				\bar{h} = \frac{1}{R_\mathrm{RH}} \sum_{J=1}^{\infty} G_J \mathds{E} [L_J^{-}].
			\end{equation}}}
			For the given $J$, if $J \geq C_\mathrm{m}+1$, the average remaining map size of Eq.~(\ref{eq_remain_map_size}) is
			\begin{equation}
				\small
				\label{equation_L_remain_def}
				{\mathds{E}} [L_J^{-} | J \geq C_\mathrm{m}+1]
				= 0 \cdot P_\mathrm{acc} + \int\limits_{0}\limits^{\frac{L_\mathrm{m}}{R_\mathrm{HM}}} \left( L_\mathrm{m} -t R_\mathrm{HM} \right) P_{T_J^{+}} (t) \mathrm{d} t,
			\end{equation}
			where the first term corresponds the vehicles which have cached the whole map before entering $B_J$.
			Substituting Eq.~(\ref{eq_T_pdf}) into Eq.~(\ref{equation_L_remain_def}), we obtain
			\begin{subequations}
				\label{eq_L_J_1}
				\small
				\begin{align}
					& {\mathds{E}} [L_J^{-} | J \geq C_\mathrm{m}+1]  \nonumber\\
					& = L_\mathrm{m}(1-P_{\mathrm{acc},J}) - R_\mathrm{HM} \int\limits_{0}\limits^{\frac{L_\mathrm{m}}{R_\mathrm{HM}}} t \cdot \frac{{\mu_\mathrm{v}}^{KC_\mathrm{m}} t^{K C_\mathrm{m}-1}}{(KC_\mathrm{m}-1)!} e^{-\mu_\mathrm{v}t} \mathrm{d} t, \nonumber\\
					& = L_\mathrm{m}(1-P_{\mathrm{acc},J}) - \frac{R_\mathrm{HM}}{\mu_\mathrm{v}} \int\limits_{0}\limits^{\frac{L_\mathrm{m}}{R_\mathrm{HM}}} \frac{{\mu_\mathrm{v}}^{KC_\mathrm{m}+1} t^{K C_\mathrm{m}}}{(KC_\mathrm{m}-1)!} e^{-\mu_\mathrm{v}t} \mathrm{d} t, \nonumber\\
					& = L_\mathrm{m} \frac{\gamma(KC_\mathrm{m},x)}{(KC_\mathrm{m}-1)!} -  \frac{R_\mathrm{HM}}{\mu_\mathrm{v}} \frac{\gamma(KC_\mathrm{m}+1,x)}{(KC_\mathrm{m}-1)!} \\
					& = L_\mathrm{m} \left[ \frac{\gamma(KC_\mathrm{m},x)}{(KC_\mathrm{m}-1)!} -  \frac{KC_\mathrm{m}\gamma(KC_\mathrm{m},x)+ x^{KC_\mathrm{m}} e^{-x}}{x(KC_\mathrm{m}-1)!} \right] \\
					& = L_\mathrm{m} \left[ (1-\frac{KC_\mathrm{m}}{x}) \frac{\gamma(KC_\mathrm{m},x)}{(KC_\mathrm{m}-1)!} + \frac{x^{KC_\mathrm{m}-1} e^{-x}}{(KC_\mathrm{m}-1)!} \right] \nonumber ,
				\end{align}
			\end{subequations}	
			{{where (\ref{eq_L_J_1}a) is according to the definition of Eq.~(\ref{eq_gamma_function}), and (\ref{eq_L_J_1}b) is based on the property of lower incomplete gamma function.}}
			Similarly, 
			\begin{equation}
				\label{eq_L_J_2}
				\small
				\begin{split}
				{\mathds{E}} [L_J^{-} | J \leq C_\mathrm{m}] & =  L_\mathrm{m} \left[ (1-\frac{K(J-1)}{x}) \frac{\gamma(K(J-1),x)}{(K(J-1)-1)!} \right. \\
				& \left. + \frac{x^{K(J-1)-1} e^{-x}}{(K(J-1)-1)!} \right].
				\end{split}
			\end{equation}
			Combining Eqs.~(\ref{eq_L_J_1}) and (\ref{eq_L_J_2}), the average service time of the RSUs is obtained:
			\begin{equation}
				\small
				\begin{split}
					& \bar{h} = \frac{1}{R_\mathrm{RM}} \sum_{J=1}^{\infty} G_J {\mathds{E}} [L_J^{-}]  = \frac{L_\mathrm{m}}{R_\mathrm{RM}} \sum_{J=1}^{\infty} \left\{ G_1 + \sum_{J=2}^{J=C_\mathrm{m}} G_J \left[  \right.\right. \\
					& \left. \left(1-\frac{K(J-1)}{x}\right) \frac{\gamma(K(J-1),x)}{(K(J-1)-1)!} +  \frac{x^{K(J-1)-1} e^{-x}}{(K(J-1)-1)!} \right] +  \\
					& \left. \left(\sum_{J=C_\mathrm{m}+1}^{\infty} G_J \right) \left[ \left(1-\frac{KC_\mathrm{m}}{x}\right) \frac{\gamma(KC_\mathrm{m},x)}{(KC_\mathrm{m}-1)!} + \frac{x^{KC_\mathrm{m}-1} e^{-x}}{(KC_\mathrm{m}-1)!} \right] \right\}				
				\end{split}
			\end{equation}	
			Apparently, the average service time increases linearly with the map size $L_\mathrm{m}$ and the inverse of RSU transmission rate $R_\mathrm{RM}$. 
			Furthermore, the average service time is proved to depend on the HAP broadcast resource and vehicle mobility, demonstrated as Proposition~2.
			
			\textbf{Proposition~2.} The average service rate of the RSU for remaining map downloading increases with the HAP broadcast rate $R_\mathrm{HM}$ and decreases with vehicle speed $1/\mu_\mathrm{v}$.
			
			\emph{Proof.} Please refer to Appendix~\ref{proof_proposition_2}.
			\hfill \rule{4pt}{8pt}\\
			
			The variance of service time is given by
			\begin{equation}
				\label{eq_variance}
				\bar{h^2} - (\bar{h})^2 = \frac{1}{{R_\mathrm{RM}}^2}\sum_{j=1}^{\infty} \mathds{E} [{L_J^{-}}^2] - ({\mathds{E}[{L}_J^{-}]})^2,
			\end{equation}
			where $\bar{h^2}$ is the mean square of service time, and $\mathds{E} [{L_J^{-}}^2]$ can be analyzed in the same way as $\mathds{E}[{L}_J^{-}]$.
			For the given $J\geq C_\mathrm{m}+1$,
			\begin{equation}
				\label{eq_L_J_squ}
				\begin{split}
					& \mathds{E} [{L_J^{-}}^2] =  \int\limits_{0}\limits^{\frac{L_\mathrm{m}}{R_\mathrm{HM}}} \left( L_\mathrm{m} - t R_\mathrm{HM}  \right)^2 P_{T_J^{+}}(t) \mathrm{d} t \\
					& = {L_\mathrm{m}}^2 \left[ \frac{\gamma(KC_\mathrm{m}-1)}{(KC_\mathrm{m}-1)!} - 2 \frac{\gamma(KC_\mathrm{m}+1,x)}{x(KC_\mathrm{m}-1)!} + \frac{\gamma(KC_\mathrm{m}+2,x)}{x^2 (KC_\mathrm{m}-1)!}  \right]
				\end{split}
			\end{equation}
			For the given $J\leq C_\mathrm{m}$, $\mathds{E} [{L_J^{-}}^2]$ can be obtained by replacing $C_\mathrm{m}$ with $J-1$ in (\ref{eq_L_J_squ}).
			Then, the close form of (\ref{eq_variance}) can be obtained.
			
			{{According to the Pollaczek--Khinchine formula which states the relationship between the queue length and service time distribution Laplace transform for an M/G/1 queue \cite{PK_formula}, the average waiting time of an M/G/1 system is given by}}
			\begin{equation}
				\bar{W}_\mathrm{M} = \bar{h} + \frac{ \lambda_\mathrm{v} \bar{h^2} }{2(1-\lambda_\mathrm{v}\bar{h})},
			\end{equation}
			where the first term denotes the service time and the second term is the queueing time.
			By substituting $\bar{h}$ and $\bar{h^2}$, the total delay of map downloading at the RSU can be obtained.
			
			\subsection{Delay-Constrained Resource Trading of MaNa Slice}
			
			The M/G/1 model based analysis can provide the numerical results for the given system parameters.
			Furthermore, we introduce M/D/1 based approximated analysis to derive the upper bound of the average map downloading time. 
			{{The arrival rate of the M/D/1 queue is set as $\lambda'_\mathrm{v} = \lambda_\mathrm{v} (1-P_\mathrm{acc})$, i.e., the RSUs only serve vehicles without the complete map in cache. 
					The service time is set as $\hat{h} = \frac{L_\mathrm{m}}{R_\mathrm{RM}}$.
					As the service time in the M/G/1 queue is no larger than $\hat{h}$, the M/D/1 model corresponds to the case where all vehicles without the complete map need to download the whole map file from the RSU.
					Therefore, the M/D/1 based analysis is a conservative approximation of the reality.}}
			
			According to the M/D/1 queueing model, the map downloading delay is given by
			\begin{equation}
				\label{eq_waiting_time_18}
				\begin{split}
					& \hat{W}_\mathrm{M} = \hat{h} + \frac{\lambda'_\mathrm{v} \hat{h}^2}{2\left( 1-\lambda'_\mathrm{v} \hat{h} \right)}
					= \frac{L_\mathrm{m}}{R_\mathrm{RM}} + \frac{\lambda'_\mathrm{v} \left(\frac{L_\mathrm{m}}{R_\mathrm{RM}}\right)^2}{2\left( 1-\frac{L_\mathrm{m}}{R_\mathrm{RM}}\lambda'_\mathrm{v}\right)}.
				\end{split}
			\end{equation}	
			Then, the constraint (\ref{eq_P1}c) of the problem (P1) can be strengthened as $\hat{W}_\mathrm{M} \leq \hat{T}_\mathrm{m}$.
			Furthermore, the RSU transmission rate requirement $R_\mathrm{RM}$ can be derived in closed form, given by Proposition~3.
			
			\textbf{Proposition~3.} The average map downloading delay at RSUs can be guaranteed if the RSU transmission rate $R_\mathrm{RM}$ satisfies:
			\begin{equation}
				\label{eq_R_RM_result}
				R_\mathrm{RM} \geq \frac{\lambda'_\mathrm{v} \hat{T}_\mathrm{m}}{1+\lambda'_\mathrm{v} \hat{T}_\mathrm{m} -\sqrt{(\lambda'_\mathrm{v}\hat{T}_\mathrm{m})^2+1}} \frac{L_\mathrm{m}}{\hat{T}_\mathrm{m}},
			\end{equation}
			where $\lambda'_\mathrm{v} = \lambda_\mathrm{v} (1-P_\mathrm{acc})$ denoting the arrival rate of vehicles without complete map in cache.
			
			\emph{Proof.} Please refer to Appendix~\ref{proof_proposition_3}.
			\hfill \rule{4pt}{8pt}\\
			
			Proposition~3 reveals the resource provisioning for the MaNa slice.
			For the given HAP broadcast rate and vehicle cache size, the accomplishment ratio $P_\mathrm{acc}$ can be obtained with Eqs.~(\ref{eq_P_acc_J}) and (\ref{eq_P_acc}).
			Substituting the result into Eq.~(\ref{eq_R_RM_result}), we obtain how much RSU transmission rate is required.
			More importantly, Proposition~3 also explains the 3D resource trading relationship among the RSU transmission rate, HAP broadcast rate, and vehicle cache size.
			By setting $P_\mathrm{acc}=0$, Eq.~(\ref{eq_R_RM_result}) provides the RSU transmission rate demand for the MaNa slice without HAP proactive pushing or vehicle caching.
			As the rate $R_\mathrm{RM}$ decreases with $P_\mathrm{acc}$, Eq.~(\ref{eq_R_RM_result}) in fact reflects how much RSU transmission rate can be saved with respect to HAP broadcast rate and vehicle cache size.
			
			\textbf{Proposition~4.}~The RSU transmission rate requirement of the MaNa slice is a convex increasing function of $(1-P_\mathrm{acc})\lambda_\mathrm{v}\hat{T}_\mathrm{m}$, i.e., the normalized remaining traffic load at the RSU with HAP-vehicle proactive pushing.
			
			\emph{Proof.} Please refer to Appendix~\ref{proof_proposition_4}.
			\hfill \rule{4pt}{8pt}\\
			
			{{Although the RSU transmission rate can be reduced by increasing accomplishment ratio, the marginal gain will diminish with $P_\mathrm{acc}$ due to the convexity. }}
			Consider an extreme case of $P_\mathrm{acc}\rightarrow 1$, where almost all vehicles can download the complete map from the HAP and the RSU is extremely lightly loaded.
			The queueing delay approaches to zero, and vehicle only experience the service delay.
			Yet, the RSUs still need to provision resources with $R_\mathrm{RM} > \frac{L_\mathrm{m}}{\hat{T}_\mathrm{m}}$.
			Therefore, the multi-resources should be matched appropriately at the MaNa slice for efficient utilization.
			
\section{FoCI Slice Analysis and Provisioning}
    \label{sec_file_analysis}
    This section focuses on the resource provisioning of the FoCI slice.
    Specifically, the on-board content hit ratio is analyzed to investigate the effectiveness of HAP-vehicle pushing in the FoCI slice.
    In addition, the RSU transmission rate saving is obtained for the given delay requirement.
    
    \subsection{On-board Hit Ratio Analysis}
    
    {{In the FoCI slice, the status of vehicle cache is illustrated in Fig.~\ref{fig_FoCI_status}.
    		Denote by $i$ the number of valid files stored on-board, where $i=0,1,\cdots,C_\mathrm{p}$.}}
    State-$i$ transits to state-$(i+1)$ when the vehicle downloads a new file into the cache, and transits to state-$(i-1)$ when a file expires.	
    The system steady-state probability $[r_0,r_1,\cdots,r_i,\cdots,r_{C_\mathrm{p}}]$ can be derived based on the condition:
    \begin{equation}
    	\label{eq_popular_steady_condition}
    	\frac{R_\mathrm{HP}}{L_\mathrm{p}} r_i = \mu_\mathrm{p} r_{i+1},
    \end{equation}
    for $i=0,1,...,C_\mathrm{p}-1$.	
    Denote by $\rho = \frac{\mu_\mathrm{p} L_\mathrm{p}}{R_\mathrm{HP}}$ the inverse of normalized file update rate.
    $\rho<1$ indicates constrained vehicle cache resource, corresponding to the case that the file update rate is larger than the file expire rate.
    In contrast, $\rho>1$ indicates that the HAP broadcast rate is constrained, which may degrade the cache utilization.
    $\rho=1$ achieves rate balance.
    Therefore, the normalized file update rate reflects the resource matching status.	
    By substituting Eq.~(\ref{eq_popular_steady_condition}) into $\sum_{i=0}^{C_\mathrm{p}} = 1$, the steady-state probability can be obtained
    \begin{equation}
    	\label{eq_popular_steady_probability}
    	r_i = \left\{ \begin{array}{ll}
    		\frac{1-\rho}{1-\rho^{C_\mathrm{P}+1}} \rho ^{C_\mathrm{p}-i}, & ~~\mbox{if}~~ \rho<1,\\			
    		\frac{1-{\rho}^{-1}}{1-\rho^{-C_\mathrm{P}-1}} \left( \frac{1}{\rho} \right)^{i}, & ~~\mbox{if}~~\rho>1,\\
    		\frac{1}{C_\mathrm{p}+1}, & ~~\mbox{if}~~\rho=1,
    	\end{array} \right.
    \end{equation}
    where $r_i$ has the same form in cases of $\rho>1$ and $\rho<1$.
    Thus, the content hit ratio is:
    \begin{equation}
    	\label{eq_popular_hit_rate}
    	\begin{split}
    		& P_\mathrm{hit}= \sum_{i=1}^{C_\mathrm{p}} r_i \left(\sum_{f=1}^{i} p_f \right) = \sum_{i=1}^{C_\mathrm{p}} p_f \sum_{i=f}^{C_\mathrm{p}} r_i 
    		\\
    		& = 
    		\left\{ \begin{array}{ll}
    			\sum_{f=1}^{C_\mathrm{p}} p_f \frac{1-\rho^{C_\mathrm{P}-f+1}} {1-\rho^{C_\mathrm{P}+1}} , & ~~\mbox{if}~~\rho\neq 1,\\
    			\sum_{f=1}^{C_\mathrm{p}} p_f \frac{C_\mathrm{p}-f+1}{C_\mathrm{p}+1}, & ~~\mbox{if}~~\rho=1.	
    		\end{array}\right.		
    	\end{split}
    \end{equation}
    According to (\ref{eq_popular_hit_rate}), the content hit ratio mainly depends on the normalized file update rate, and the influences of system parameters are summarized in Propositions~5 and 6.
    
    \textbf{Proposition~5.} The on-board content hit ratio increases with the vehicle cache size for the given normalized file update rate.
    
    \emph{Proof.} 
    Notice that
    \begin{equation}
    	\begin{split}
    		& \frac{1-\rho^{a+1}}{1-\rho^{A+1}} - \frac{1-\rho^{a}}{1-\rho^{A}}
    		\frac{(1-\rho)(\rho^a-\rho^A)}{(1-\rho^A)(1-\rho^a)} \geq 0, ~~\mbox{for}~~ \rho\neq 1,				
    	\end{split}
    \end{equation}
    and 
    \begin{equation}
    	\frac{C_\mathrm{p}-f+1}{C_\mathrm{p}+1} = 1-\frac{f}{C_\mathrm{p}+1},
    \end{equation}
    increases with $C_\mathrm{p}$.
    Therefore, $P_\mathrm{hit}$ increases with $C_\mathrm{p}$ for the given $\rho$.
    \hfill \rule{4pt}{8pt}\\
    
    \textbf{Proposition~6.} The on-board content hit ratio increases with the normalized file update rate.
    
    \emph{Proof.} Please refer to Appendix~\ref{proof_proposition_6}.
    \hfill \rule{4pt}{8pt}\\
    
    {{Propositions 5 and 6 indicate how to increase the on-board content hit ratio through HAP proactive pushing in the FoCI slice. Specifically, the vehicle cache size and the normalized file update rate are the two critical parameters which directly determine the HAP offloading capability. Accordingly, the insights are two-fold: (1) sufficient vehicle cache resource should be provisioned to guarantee the on-board content hit ratio, for the given HAP file update rate; and (2) the HAP broadcast rate should adapt to the file life time, so as to keep the files cached on-board fresh.}}
    
	\begin{table}[!t]
		\caption{Simulation parameters}
		\label{tab_parameter}
		\centering
		\begin{tabular}{cccc}
			\hline
			\hline
			Parameter & Value & Parameter & Value \\
			\hline
			$K$ & 5 & $\mu_\mathrm{v}$ & 0.2 /s \\
			$L_\mathrm{m}$ & 5 Gb & $L_\mathrm{p}$ & 1 Gb \\
			$\lambda_\mathrm{v}$ & 1.2 /s & $\lambda_\mathrm{p}$ & 4 /s \\
			$F$ & 1000 & $\nu$ & 0.56 \\
			$L_\mathrm{v}$ & 200 Gbps & $R_\mathrm{R}$ & 10 Gpbs \\ 	
			\hline
			\hline
		\end{tabular}
	\end{table}

    \subsection{RSU Delay Analysis of FoCI Slice}
    
    Denote by $\lambda_\mathrm{p}$ the arrival rate of the FoCI slice, among which $P_\mathrm{hit} \lambda_\mathrm{p}$ requests can be self-served on board.
    The remaining $\lambda'_\mathrm{p} = (1-P_\mathrm{hit}) \lambda_\mathrm{p}$ requests are served by the RSU, which can be modeled as an M/D/1 queueing system at service rate of $R_\mathrm{RP}/L_\mathrm{p}$.
    The average delay of FoCI is given by
    \begin{equation}
    	\bar{W}_\mathrm{P} = \frac{L_\mathrm{p}}{R_\mathrm{RP}} + \frac{\lambda'_\mathrm{p} \left( \frac{L_\mathrm{p}}{R_\mathrm{RP}} \right)^2}{2\left( 1- \lambda'_\mathrm{p} \frac{L_\mathrm{p}}{R_\mathrm{RP}} \right)}.
    \end{equation}
    The item $\frac{\lambda'_\mathrm{p} L_\mathrm{p}}{R_\mathrm{RP} }$ is the normalized RSU load, which should be smaller than 1 for system stability.
    
    \textbf{Proposition 7.} The delay-constrained RSU transmission rate in the FoCI slice is given by
    \begin{equation}
    	\label{eq_R_RP_req}
    	R_\mathrm{RP} \geq \frac{L_\mathrm{p} \lambda'_\mathrm{p} }{\hat{T}_\mathrm{p} \lambda'_\mathrm{p} +1 - \sqrt{(\hat{T}_\mathrm{p} \lambda'_\mathrm{p})^2+1}},
    \end{equation}
    where $\hat{T}_\mathrm{p}$ is the required delay, and $\lambda'_\mathrm{p} = \lambda_\mathrm{p} (1-P_\mathrm{hit})$ is the downloading request arrival rate at each RSU.
    
    \emph{Proof.} Proposition~7 can be proved in a similar way of Proposition~3.
    \hfill \rule{4pt}{8pt}\\
    
    \textbf{Remark~3.} Proposition~7 demonstrates the 3D resource trading relationship in the FoCI slice.
    
    According to Eq.~(\ref{eq_R_RP_req}), $R_\mathrm{RP}$ can be proved to decrease with $P_\mathrm{hit}$.
    As the hit ratio $P_\mathrm{hit}$ has been proved to increase with $C_\mathrm{p}$ and $R_\mathrm{HP}$, the required minimal RSU transmission rate decreases with vehicle cache size and HAP broadcast rate. 
    Substituting $P_\mathrm{hit}=0$ (${C_\mathrm{p}}=0,~R_\mathrm{HP}=0$) into Eq.~(\ref{eq_R_RP_req}), we obtain the minimal RSU transmission rate without HAP-vehicle proactive pushing.
    As $C_\mathrm{p}$ and $R_\mathrm{HP}$ increase, the variation of $R_\mathrm{RP}$ in Eq.~(\ref{eq_R_RP_req}) indicates how much RSU transmission rate can be traded with HAP broadcast and vehicle cache resources.
    
    \begin{figure*}[!t]
    	\centering
    	\subfloat[] {\includegraphics[width=2.5in]{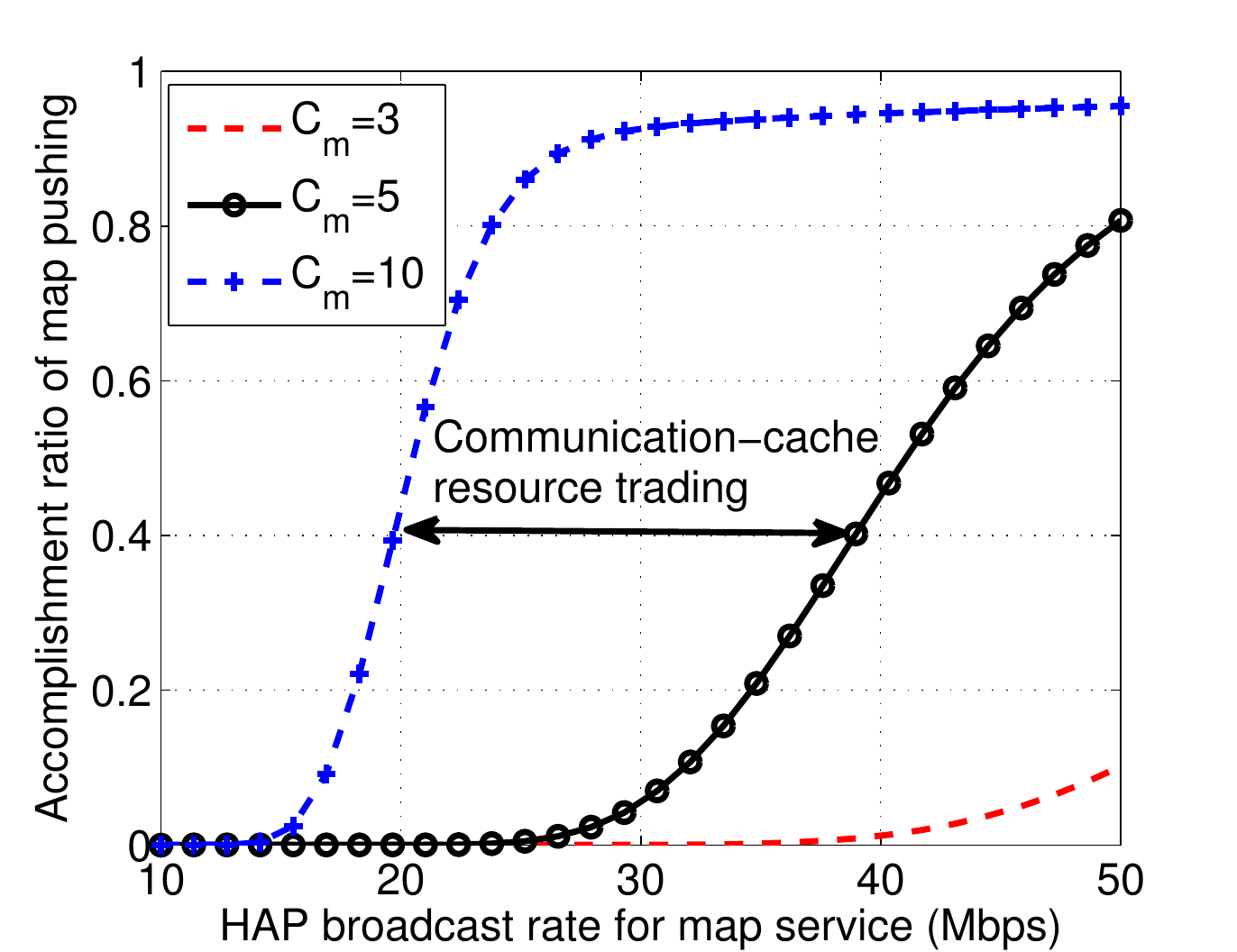}}
    	\hfil
    	\subfloat[]{\includegraphics[width=2.5in]{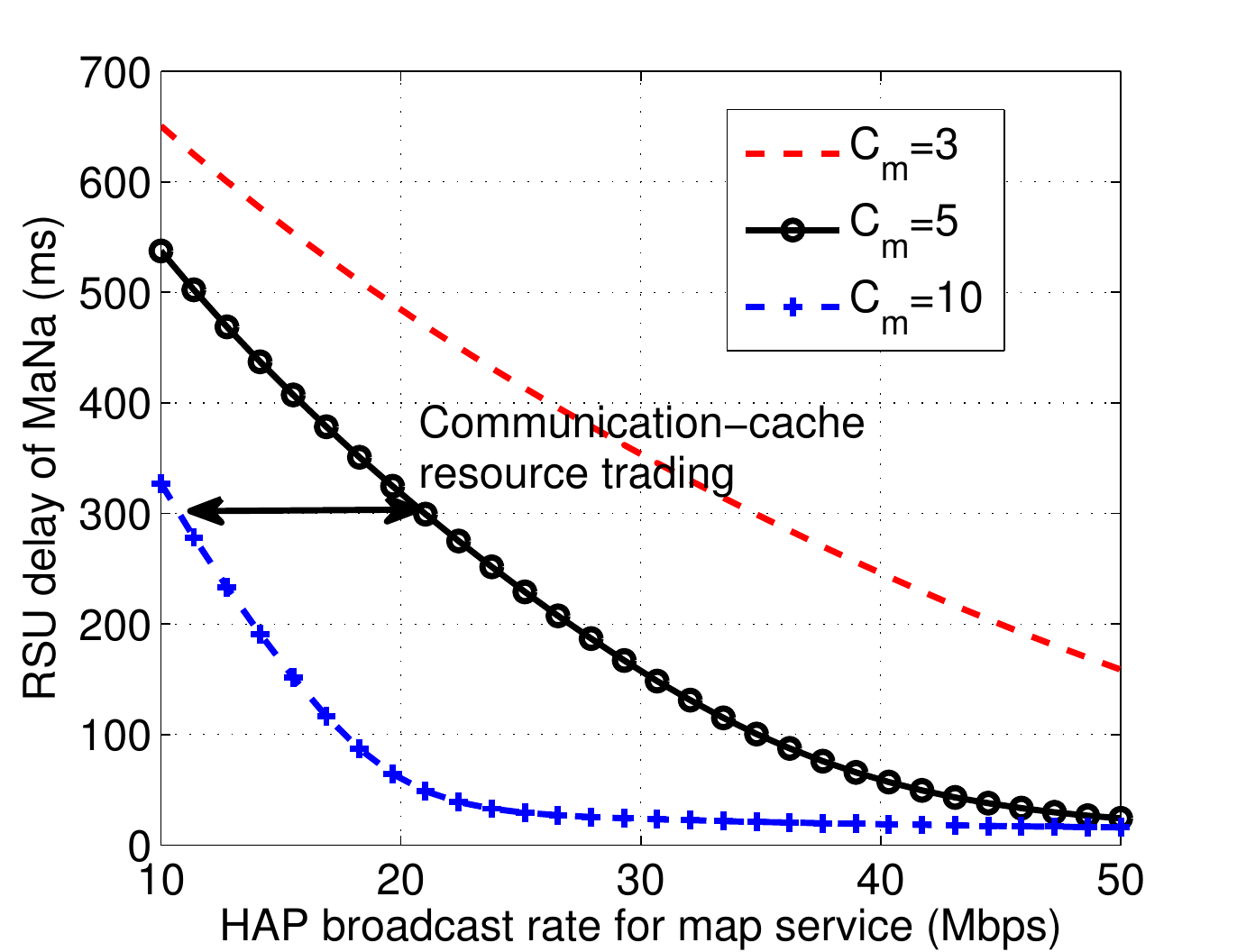}}
    	\caption{MaNa performance improvement with HAP-vehicle proactive pushing and caching, (a) accomplishment ratio of HAP-vehicle map pushing, (b) average delay at RSU, $C_m$ denotes the vehicle cache size normalized by the map file size.}
    	\label{fig_map_2D}
    \end{figure*}
    
\section{Simulation and Numerical Results}
    \label{sec_simulation}
    This section provides numerical results to validate the analytical results of the MaNa slice and FoCI slice, from the aspects of HAP offloading capability, multi-resource trading, and important system parameters. 
    In addition, the network performance is illustrated under different slicing schemes, demonstrating multi-resource matching across slices in dynamic resource sharing.
    
    The simulated scenario consists of one HAP cell covering 10 four-lane blocks, where each block is of length 1 km and generates an HD map of size 5 Gb. 
    The large scale vehicle mobility is considered to be memoryless. 
    The vehicles are considered to arrive destination with probability $\psi$, and continue to next block with probability $1-\psi$ in each block.
    The dwelling time of a vehicle on each block follows Erlang distribution with $K=5$ and $\mu_\mathrm{v}=0.2$ /s, corresponding to an average speed of $40$ km/h.
    Zipf distribution is adopted for the content popularity distribution in the FoCI slice \cite{video_popularity_2009}:
    \begin{equation}
    	p_f = \frac{1/f^\nu}{\sum_{u=1}^{F} 1/u^\nu},
    \end{equation}
    where $\nu\geq 0$ is a constant parameter indicating the skewness of popularity distribution. 
    {{Parameter $\nu$ can be set as 0.56 to depict the video streaming services \cite{video_popularity_2009}, which can be even higher for the driving-related mobile applications considering the similarity of location-related requests.
    		Important parameters are listed in Table~\ref{tab_parameter} \cite{Liu16_EE_cache_JSAC}.}}

    \subsection{MaNa slice}
    
    Figure~\ref{fig_map_2D} shows the offloading capability of HAP-vehicle proactive pushing and caching in the MaNa slice.
    Fig.~\ref{fig_map_2D}~(a) demonstrates the accomplishment ratio of HAP-vehicle map downloading, i.e., how many vehicles can finish map downloading in the HAP download window.
    The simulation results reveal that the accomplishment ratio increases with HAP broadcast rate, where the increasing rate firstly increases but then decreases.
    In addition, almost all vehicles can complete map downloading before entering the corresponding block
    through the HAP map pushing, if the resources of broadcast and cache are sufficiently provisioned. 
    The results also indicate the trading relationship between HAP communication and vehicle storage resources.
    For example, to guarantee 40\% accomplishment ratio, the HAP broadcast rate is required to be 20 Mbps when each vehicle can cache 10 maps, which will be doubled when each vehicle can cache 5 maps.
    Fig.~\ref{fig_map_2D}~(b) shows the average delay at RSUs, where the RSU transmission rate is set as 10 Gpbs.
    {{Similarly, the average map downloading delay is shown to decrease with HAP broadcast rate and vehicle cache size.}}
    When the HAP broadcast rate and vehicle cache size are sufficiently high, the traffic load of RSUs approximates zero, and the file downloading delay is shown to level off and asymptotically go to zero.
    The resource trading relationship can also be found in Fig.~\ref{fig_map_2D}~(b), where the required HAP broadcast rate will significantly increase as the cache size decreases, for the given delay requirement.
    The results of Figs.~\ref{fig_map_2D}~(a) and (b) are consistent.
    As the accomplishment ratio increases, the traffic load of RSUs decreases, which can help to reduce the service delay.

    \begin{figure*}[!t]
    	\centering
    	\subfloat[] {\includegraphics[width=2.2in]{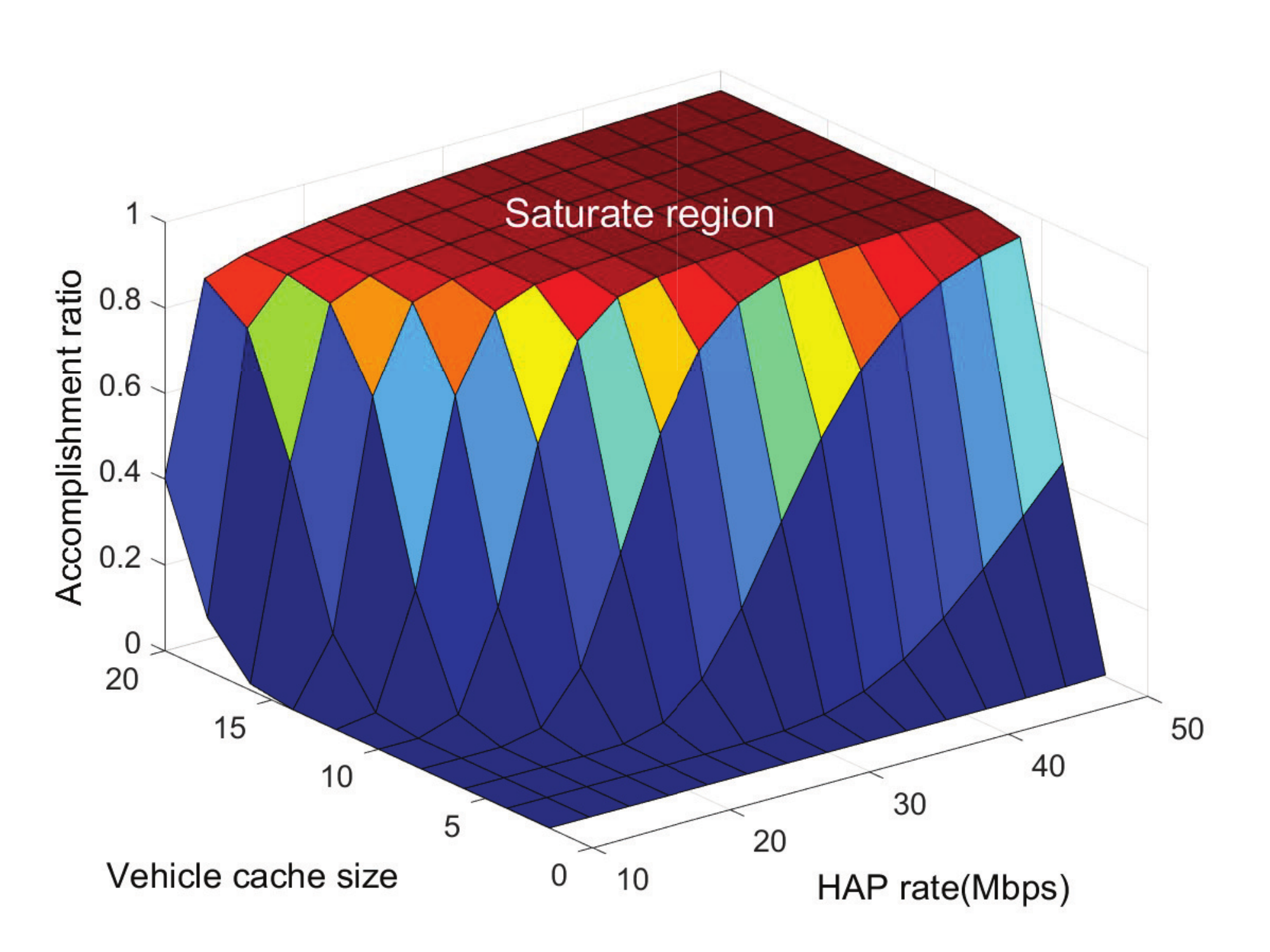}}
    	\subfloat[]{\includegraphics[width=2.2in]{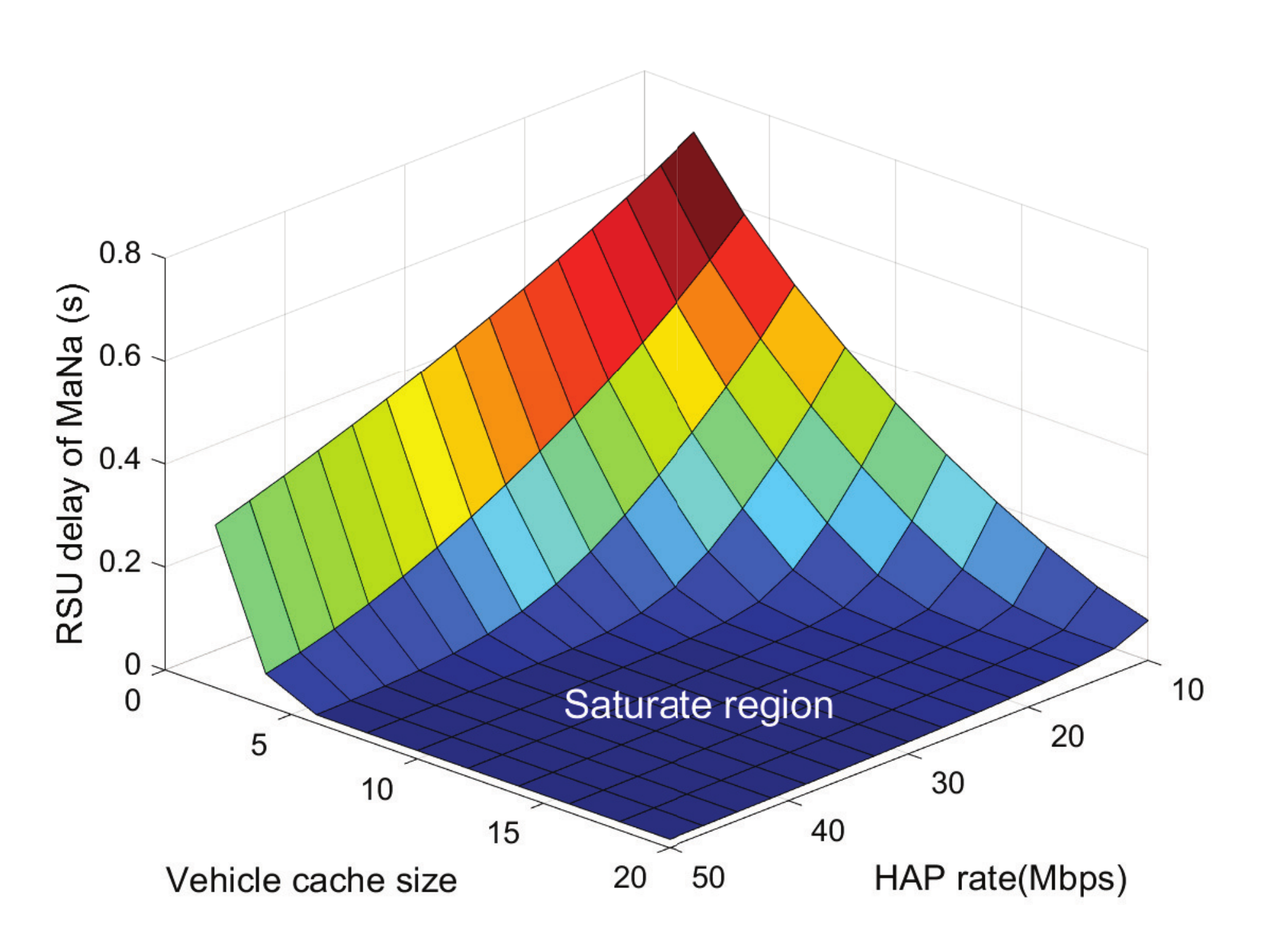}}
    	\subfloat[]{\includegraphics[width=2.2in]{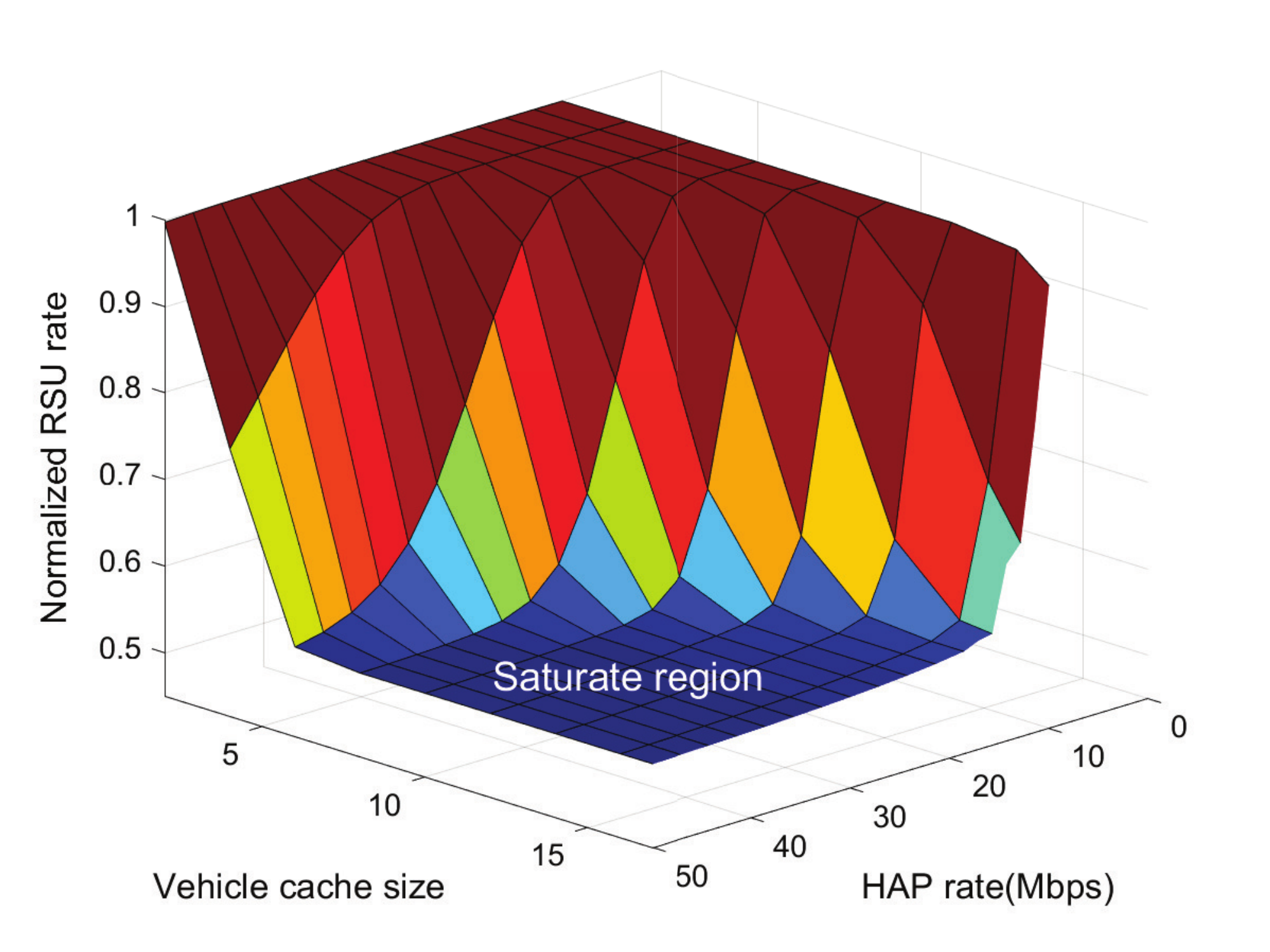}}
    	\caption{{{QoS-constrained transmission and storage resource provisioning in the MaNa slice, (a) accomplishment ratio of HAP-vehicle map pushing, (b) average delay at RSUs, (c) RSU transmission rate requirement.}}}
    	\label{fig_map_3D}
    \end{figure*}
    
    \begin{figure}[!t]
    	\centering
    	\includegraphics[width=2.5in]{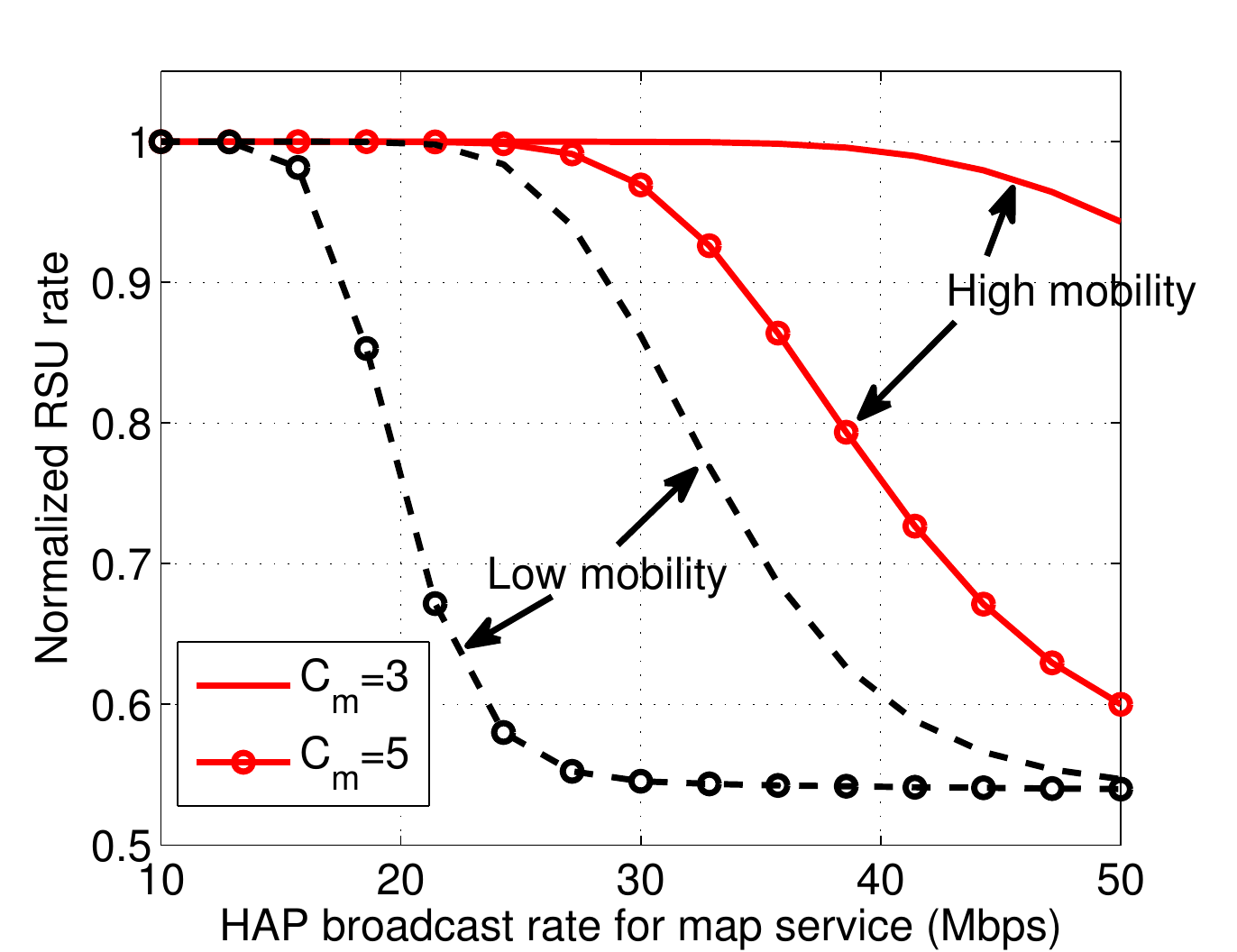}
    	\caption{Influence of vehicle mobility on MaNa slice provisioning, with the low-mobility speed of 20 km/h and high-mobility speed of 40 km/h.}
    	\label{fig_map_mobility}
    \end{figure}

    Figure~\ref{fig_map_3D} provides more details on the communication and cache resource trading from different perspectives.
    {{Fig.~\ref{fig_map_3D}~(a) shows the accomplishment ratio of HAP-assisted map downloading, and Fig.~\ref{fig_map_3D}~(b) demonstrates the average delay at RSUs, with respect to HAP broadcast rate and vehicle cache size. }}
    Fig.~\ref{fig_map_3D}~(c) shows the RSU transmission rate requirement of the MaNa slice, which is normalized by the rate without HAP-vehicle pushing or caching.
    The results of Fig.~\ref{fig_map_3D} reveal three insights.
    Firstly, the QoS performance of HAP proactive pushing can be improved by increasing either HAP broadcast rate or vehicle cache size, yet the effectivenesses are different.
    Secondly, the HAP communication resource and vehicle cache resource should be matched to effectively enhance the offloading capability of HAPs.
    For example, few vehicles can accomplish map downloading from HAPs even at high HAP broadcast rate (e.g., 50 Mbps), if the cache size is insufficient (e.g., $C_\mathrm{m}$=3).	
    The saddle-point effect also exists in Fig.~\ref{fig_map_3D}, indicating the optimal HAP and vehicle resource provisioning for the MaNa slice.
    {{Thirdly, the offloading capability of HAP-vehicle pushing cannot be further improved when the broadcast rate and cache size are sufficiently large, depicted as the saturate region.}}
    In this case, the resources can be reallocated to other slices to improve the overall network performance, through dynamic network slicing and sharing.

    \begin{figure*}[!t]
    	\centering
    	\subfloat[] {\includegraphics[width=2.5in]{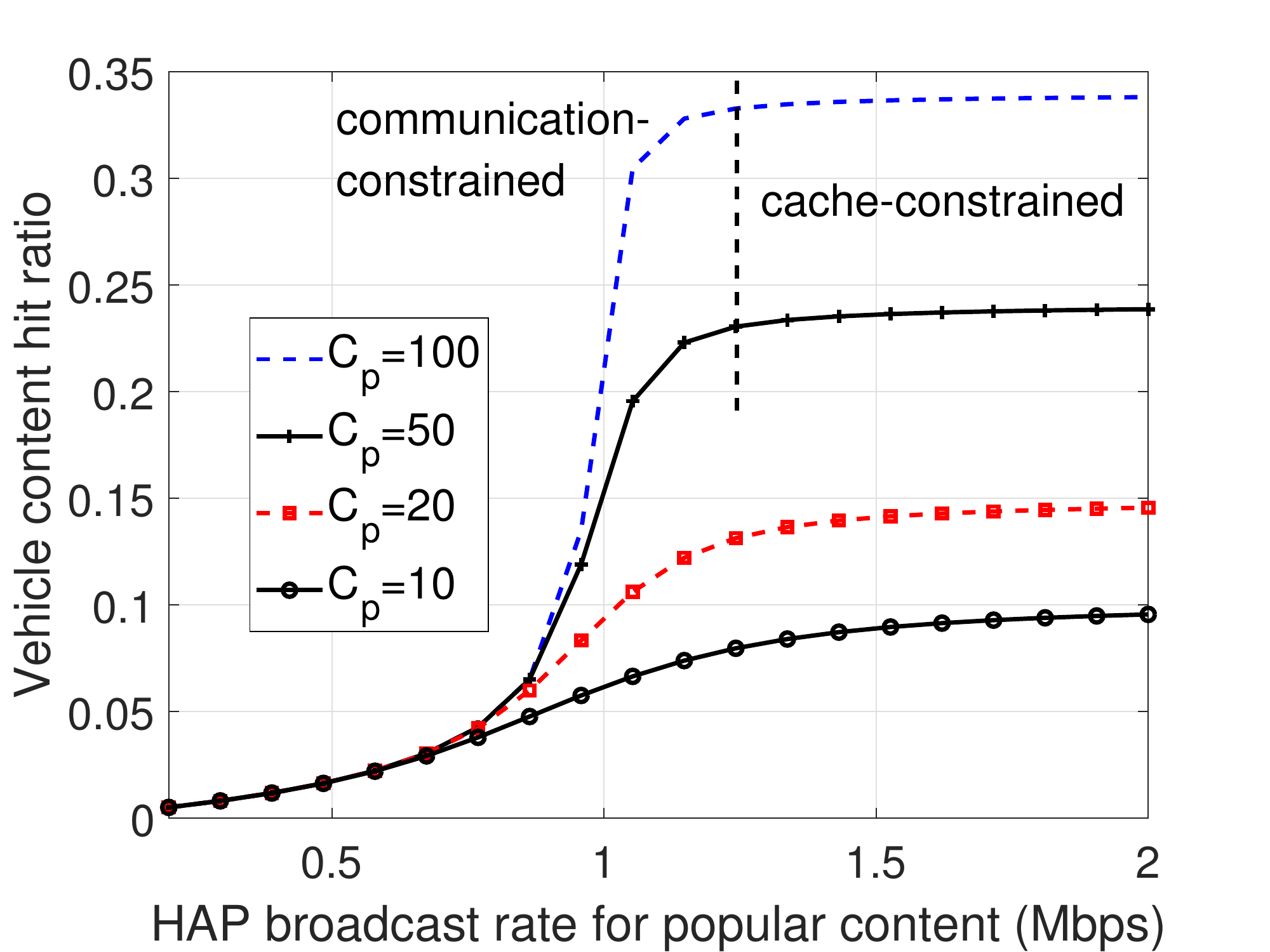}}
    	\hfil
    	\subfloat[]{\includegraphics[width=2.5in]{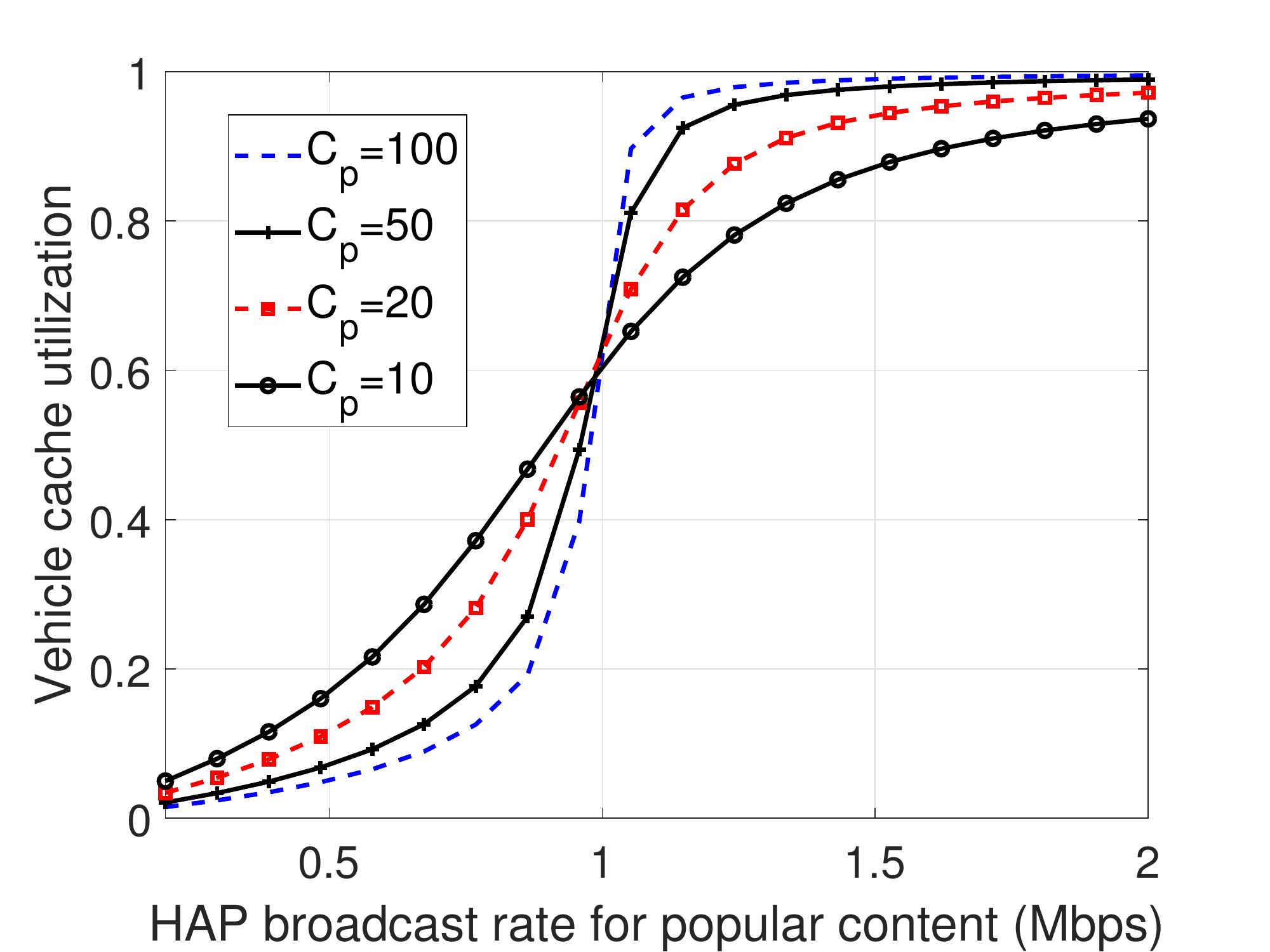}}
    	\caption{{{On-board proactive content caching efficiency of the FoCI slice, (a) vehicle content hit ratio, (b) vehicle cache utilization ratio.}}}
    	\label{fig_popular_hit_rate}
    \end{figure*}
    
    Figure~\ref{fig_map_mobility} further shows the influence of vehicle mobility on the MaNa slice performance.
    The average speed at high mobility is 40 km/h, where the parameters of Erlang dwelling time are set as $K=5$ and $\mu_\mathrm{v}=0.2$ /s.
    In comparison, the average speed at low mobility is 20 km/h, where the Erlang parameters are set as $K=10$ and $\mu_\mathrm{v}=0.2$ /s instead.
    The results reveal that high mobility of vehicles calls for additional resource provisioned in the MaNa slice. 
    The reason is that the available HAP window for a target map is shortened on average due to the high vehicle mobility, which degrades the accomplishment ratio of HAP-vehicle map pushing.
    As compensation, vehicles can start downloading the target map ahead of schedule, to enhance the accomplishment ratio.
    However, contents may overflow on-board, and real-time content caching and update schemes need to be designed to address this conflict.
    A more important insight of Fig.~\ref{fig_map_mobility} is that the resource provisioning of MaNa slice should be dynamically adjusted to vehicle mobility.
    {{For example, more caching and communication resources of the MaNa slice can be relieved and thus reused by other slices during rush hours or traffic jams. }}

    \subsection{FoCI slice}

    Figure~\ref{fig_popular_hit_rate}~(a) shows the performance of HAP proactive pushing in the FoCI slice, with respect to the HAP broadcast rate and vehicle cache size.
    Specifically, the content hit ratio firstly increases with the content hit ratio super-linearly, but then levels off around some constant.
    The turning point, around 1.1--1.2 Mbps, can be defined as \textit{saturate rate} and set as the upper bound of broadcast rate for the FoCI slice.
    Accordingly, the resource provisioning can be divided into two regions, i.e., \textit{communication constrained} and \textit{cache constrained}, as marked in Fig.~\ref{fig_popular_hit_rate}~(a).
    In the communication constrained region, the hit ratio can be significantly improved by increasing the HAP broadcast rate.
    On the contrary, in the cache constrained region, the content hit ratio can only be improved by adding more caching resources.
    Therefore, the communication and caching resources also need to be matched for the FoCI slice, considering the saturate rate.
    
    The results of Fig.~\ref{fig_popular_hit_rate}~(a) can be explained by Fig.~\ref{fig_popular_hit_rate}~(b) from the perspective of cache utilization, where the cache utilization rate denotes the ratio of valid contents to cache size.
    Specifically, the cache utilization rate increases with broadcast rate in the communication-constrained region, but levels off at around 1 in the cache-constrained region.
    When the broadcast rate is low, the cached contents cannot be timely updated, resulting in invalid cached contents and low cache utilization.
    When the broadcast rate achieves the saturate rate, the broadcast rate is sufficient to update content timely, whereby the cache can be fully utilized but the HAP broadcast resource may be underutilized.
    Therefore, the saturate rate reflects the optimal match of HAP broadcast rate and vehicle cache size for the FoCI slice resource provisioning.

    
    \begin{figure}[t]
    	\centering
    	\includegraphics[width=2.5in]{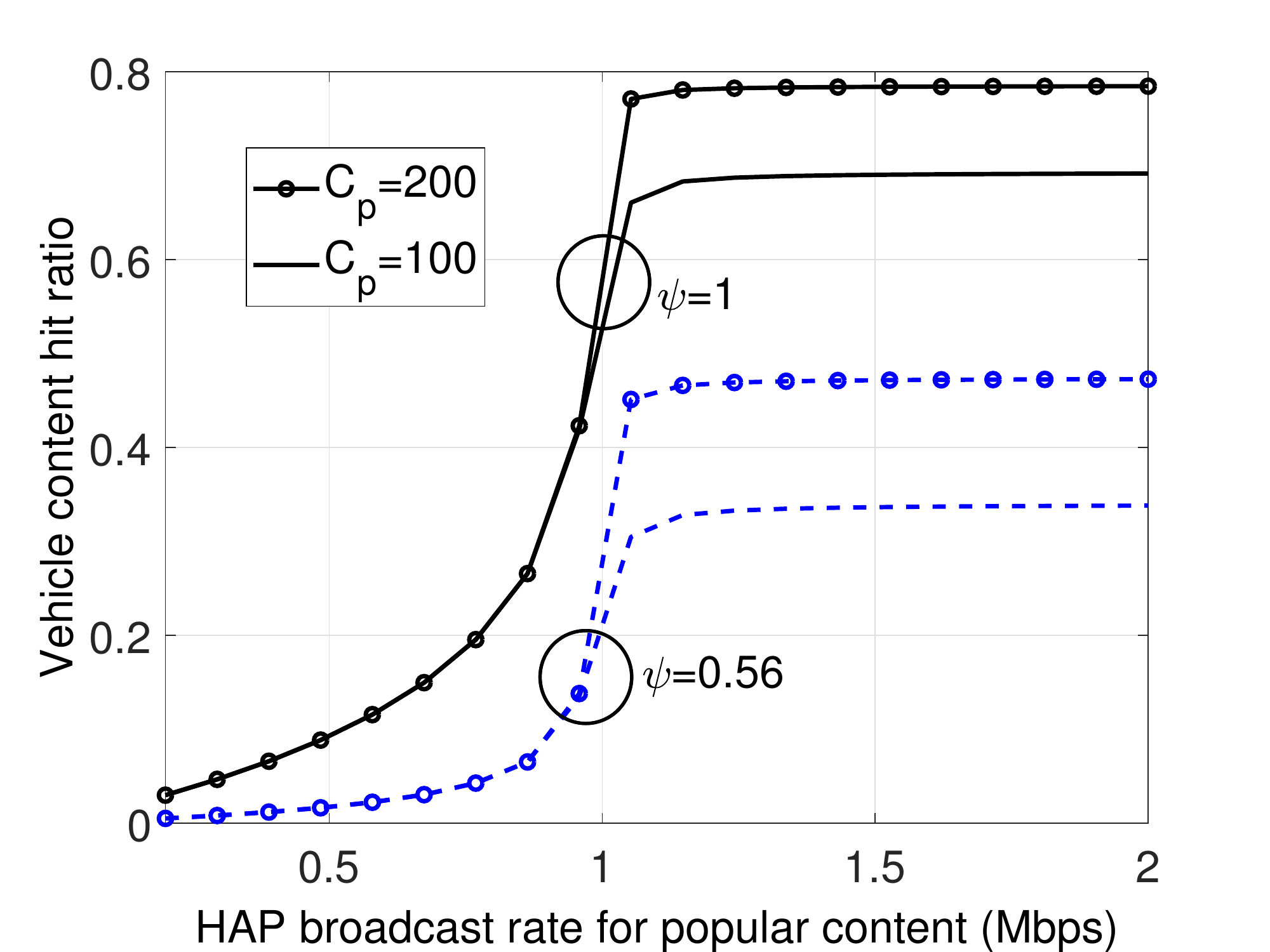}
    	\caption{Influence of content popularity distribution, $\nu$ content popularity skewness.}
    	\label{fig_popular_popularity}
    \end{figure}
    
    \begin{figure}[t]
    	\centering
    	\includegraphics[width=2.5in]{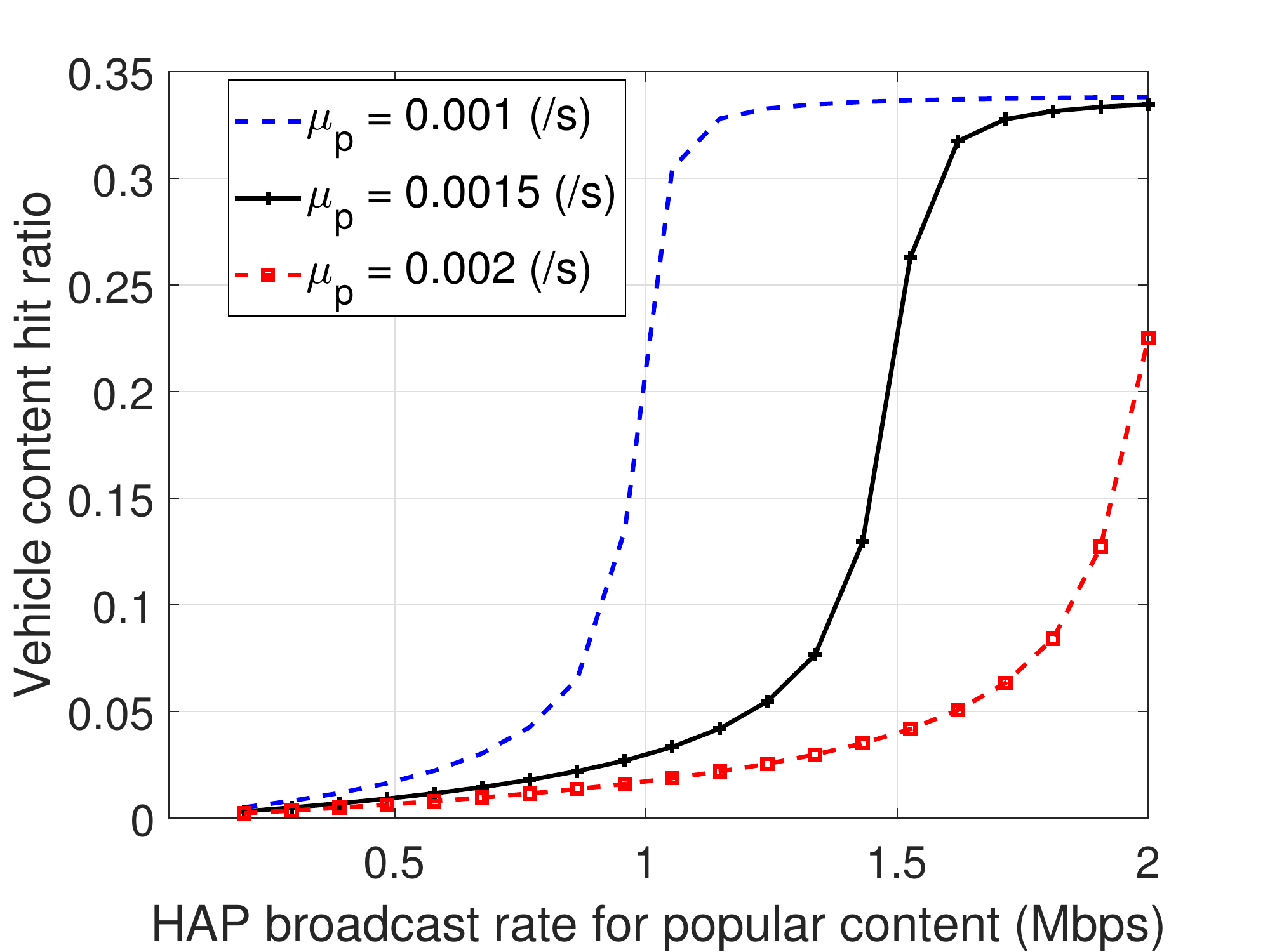}
    	\caption{Influence of file life time, $\mu_\mathrm{p}$ file expire rate.}
    	\label{fig_popular_life_time}
    \end{figure}
    
    \begin{figure*}[!t]
    	\centering
    	\subfloat[] {\includegraphics[width=2.5in]{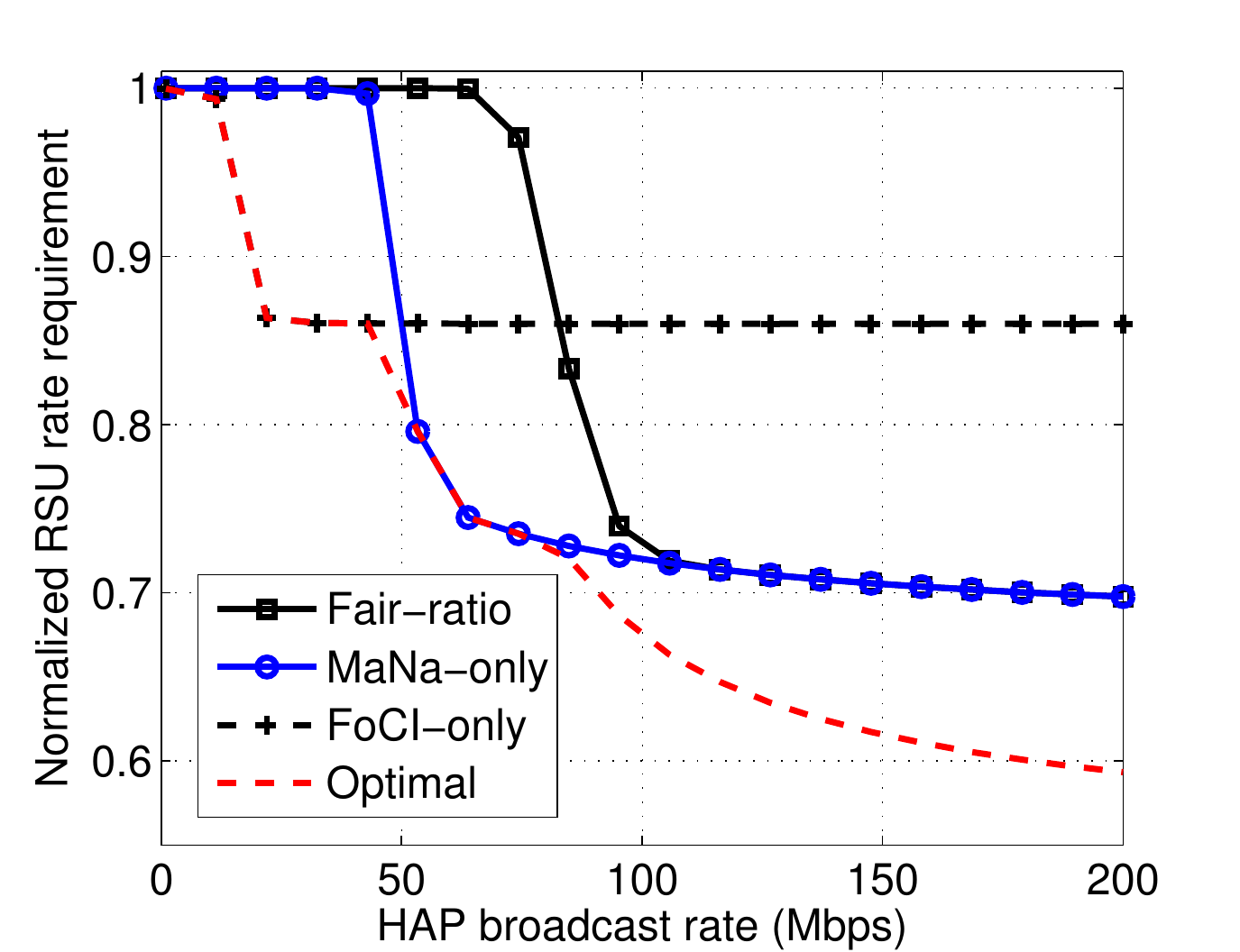}}
    	\hfil
    	\subfloat[] {\includegraphics[width=1.6in]{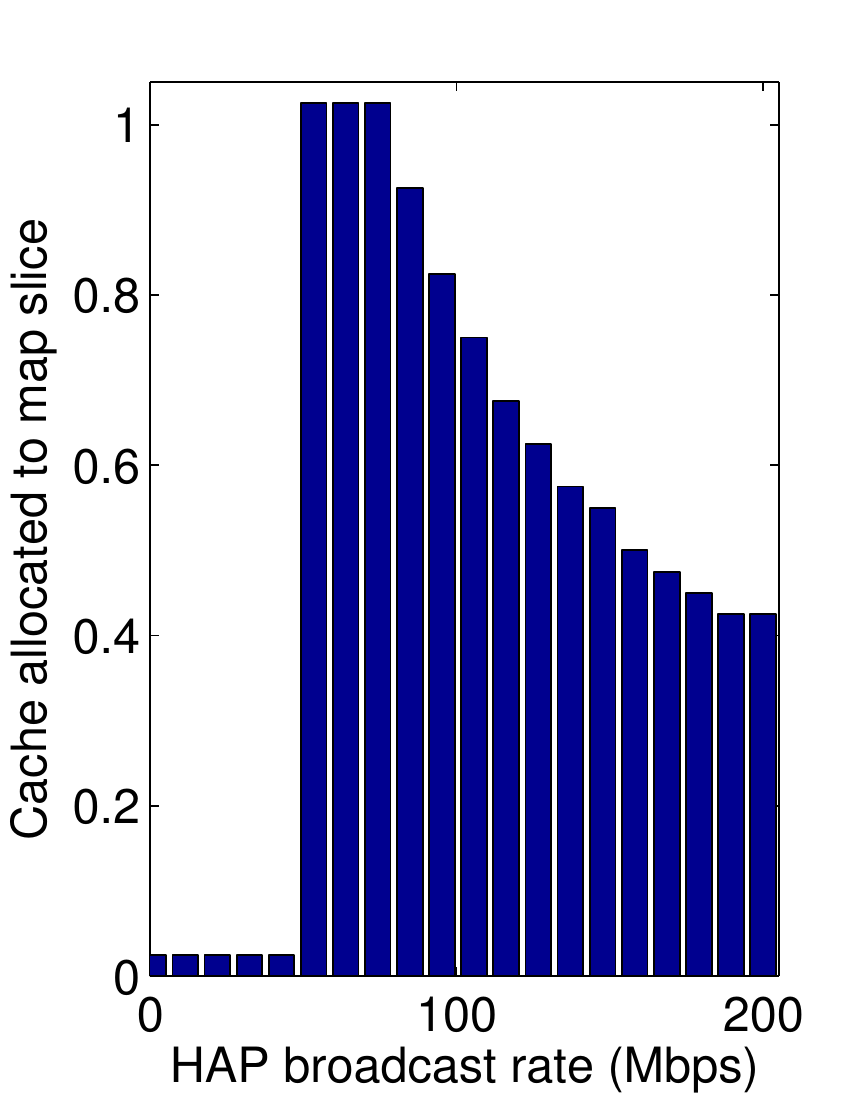}}
    	\hfil
    	\subfloat[]{\includegraphics[width=1.6in]{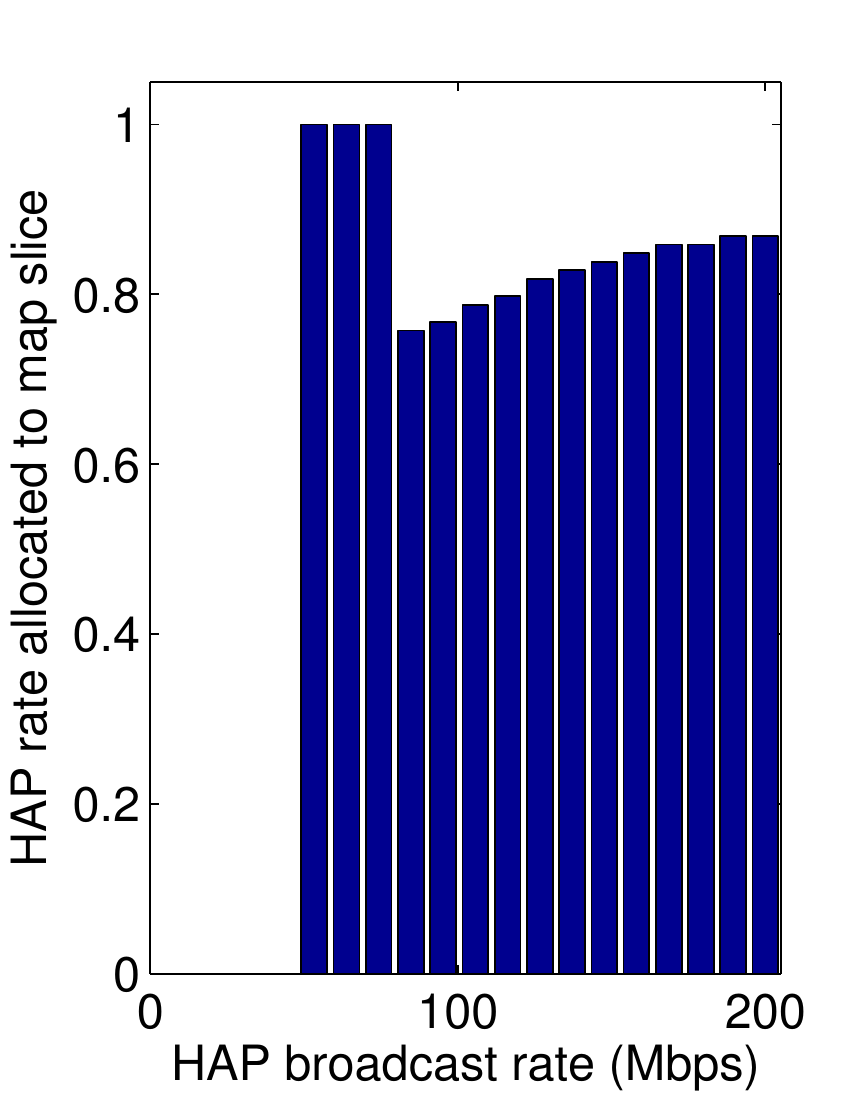}}
    	\caption{Optimal multi-resource slicing, (a) total RSU transmission rate requirement of MaNa and FoCI slices, (b) optimal vehicle cache resource allocation to the MaNa slice, (c) optimal HAP transmission rate allocation to the MaNa slice.}
    	\label{fig_slicing_scheme_compare}
    \end{figure*}
    
    The performance of proactive HAP pushing can be significantly influenced by the content request features.
    Fig.~\ref{fig_popular_popularity} demonstrates the influence of content popularity distribution on the effectiveness of proactive HAP pushing.
    The observations are as follows.
    Firstly, the HAP proactive pushing is more effective when the content requests are more concentrated (e.g., skewness factor $\nu=1$).
    This is because the cached content can gain higher hit ratio with larger $\nu$, and thus improve caching efficiency.
    Secondly, the content hit ratio gain of cache size can fade with $\nu$, since the hit ratio of less popular contents is lower if the requests concentrate on the most popular ones.
    Thirdly, saturate rate is not influenced by the skewness factor $\nu$, but varies with the file life time as shown in Fig.~\ref{fig_popular_life_time}.
    Specifically, the saturate rate increases with the file expire rate $\mu_\mathrm{p}$, indicating that the vehicle cache needs to be updated more frequently if the file life time decreases.
    The result of Fig.~\ref{fig_popular_life_time} is consistent with the analytical results, where the normalized file update rate directly determines the content hit ratio.
    
    \subsection{Network slicing and sharing}
    
    The simulation results of MaNa and FoCI slices both demonstrate the necessity of matching broadcast rate and cache size, based on network parameters and traffic features.
    Therefore, the resource provisioning for different slices needs to be jointly optimized to fully utilize the multi-dimensional heterogeneous resources.
    We study how to balance the multi-resource provisioning when MaNa and FoCI slices coexist. 
    
    The performance of different network slicing schemes are compared, as shown in Fig.~\ref{fig_slicing_scheme_compare}. The HAP broadcast rate and vehicle cache size are jointly sliced, the vehicle cache size is 200 Gb, the required downloading delays are 1 s and 5 s for the map and popular contents, respectively.
    The optimal slicing is obtained through exhaustive search, the fair-ratio scheme allocates HAP broadcast rate and vehicle cache size depending on the traffic loads of MaNa and FoCI slices, the MaNa-only scheme only enables proactive caching for the MaNa slice, and the FoCI-only scheme only enables proactive caching for FoCI slice.
    Fig.~\ref{fig_slicing_scheme_compare}~(a) shows the required RSU transmission rate, which is normalized by the one without HAP proactive pushing or vehicle caching.
    According to Fig.~\ref{fig_slicing_scheme_compare}~(a), the transmission rate demand of RSUs can be reduced by more 40\%, with HAP-vehicle proactive pushing and caching.
    However, the performance is significantly influenced by the slicing schemes.
    Specifically, the optimal matching scheme can bring around 37\% RSU transmission rate saving compared with the FoCI-only slicing scheme with HAP broadcast rate of 200 Mbps, and 10\% compared with the MaNa-only or fair-ratio schemes.
    
    The optimal slicing ratios of vehicle cache and HAP broadcast rate are demonstrated in Figs.~\ref{fig_slicing_scheme_compare}~(b) and (c), respectively.
    The optimal slicing scheme can be divided into three cases, regarding the HAP broadcast rate.
    When the broadcast rate is lower than 50 Mbps, the optimal scheme allocates all caching and HAP broadcast resources to the FoCI slice.
    In this case, the map accomplishment ratio approximates to zero when the HAP broadcast rate is low, as shown in Fig.~\ref{fig_map_2D}.
    Accordingly, the FoCI-only scheme performs better than the MaNa-only and fair ratio schemes, as shown in Fig.~\ref{fig_slicing_scheme_compare} (a).
    As the HAP broadcast rate ranges between 50--80 Mbps, the optimal scheme allocates all resources to the MaNa slice.
    The performance of popular file only scheme levels off, reflecting that the HAP broadcast rate exceeds the saturate rate.
    In comparison, the performance of MaNa-only scheme improves significantly and combats the FoCI-only scheme.
    This stage corresponds to the slope of Fig.~\ref{fig_map_2D}~(a), where the accomplishment ratio increases rapidly with the HAP broadcast rate of the MaNa slice.
    Finally, as the broadcast rate exceeds 80 Mbps, the optimal scheme enables both MaNa and FoCI slices to share the vehicle cache and HAP broadcast resources, which realizes multi-dimensional resource matching.
    
\section{Conclusions and Future Work}
    \label{sec_conclusions}

{{Under the proposed AGIVEN architecture, this paper has an in-depth investigation on the multi-dimensional heterogeneous resource provisioning and resource trading for MaNa, FoCI and ODT slicing. }}
In the MaNa slice, the trading relationship between HAP broadcast and vehicle cache size has been analyzed based on the derived accomplished ratio, which has suggested that the optimal HAP broadcast rate can decrease with the vehicle cache size in an inversely proportional manner.
In the FoCI slice, the on-board content hit ratio has been derived with respect to the normalized file update rate and cache size, where the communication-/cache-constrained conditions has indicated the optimal resource provisioning in capacity enhancement.
The offloading capabilities of HAP-vehicle pushing have been derived for the MaNa and FoCI slices, respectively, which has shown the service-dependent 3D trading relationships among HAPs, RSUs, and vehicles.
The 3D trading relationship has been applied in network slicing, whereby up to 40\% of RSU transmission rate can be saved to enhance the performance of the ODT slice.
The current work mainly targets on the large-scale slice-level multi-resource provisioning, based on the statistic information of vehicle mobility, content popularity, and traffic requests.
For the future work, we will exploit the small-time-scale information to design dynamic network slicing and sharing schemes to further enhance resource utilization. 


\appendices{}
   \section{Proof of Proposition~1}
   \label{proof_proposition_1}
   Taking the derivative of $\hat{P}_\mathrm{acc}$ with respect to $x$, we have
   \begin{equation}
   	\frac{\partial \hat{P}_\mathrm{acc}}{\partial x} = -\frac{x^{KC_\mathrm{m}-1} e^{-x}}{(KC_\mathrm{m}-1)!},
   \end{equation}
   and 
   \begin{equation}
   	\frac{\partial^2 \hat{P}_\mathrm{acc}}{\partial x^2} = - \frac{KC_\mathrm{m}-x}{(KC_\mathrm{m}-1)!}  x^{KC_\mathrm{m}-1} e^{-x}.
   \end{equation}
   As $x = \frac{L_\mathrm{m} \mu_\mathrm{v}}{R_\mathrm{HM}}$, we can derive
   \begin{equation}
   	\frac{\partial \hat{P}_\mathrm{acc}}{\partial R_\mathrm{HM}} = \frac{1}{L_\mathrm{m}\mu_\mathrm{v}} \frac{x^{KC_\mathrm{m}+2} e^{-x}}{(KC_\mathrm{m}-1)!} >0, \\
   \end{equation}
   and 
   {{\begin{equation}
   			\small
   			\begin{split}
   				\frac{\partial^2 \hat{P}_\mathrm{acc}}{\partial {R_\mathrm{HM}}^2} & = \frac{1}{L_\mathrm{m}\mu_\mathrm{v}} \frac{x^{KC_\mathrm{m}+1}e^{-x} \left[ (KC_\mathrm{m}+2)-x \right] }{(KC_\mathrm{m}-1)!} \left( -\frac{\mu_\mathrm{v} L_\mathrm{m}}{{R_\mathrm{HM}}^2} \right)\\
   				& = - \left(\frac{1}{L_\mathrm{m}\mu_\mathrm{v}}\right)^2 	\frac{x^{KC_\mathrm{m}+3}e^{-x} }{(KC_\mathrm{m}-1)!} 	\left[ (KC_\mathrm{m}+2)-x \right] \\
   				& \left\{ {\begin{array}{ll} <0, & \mbox{if}~~x < KC_\mathrm{m} +2\\ >0, & \mbox{if}~~x>KC_\mathrm{m}+2,
   					\end{array}} \right.
   				\end{split}
   			\end{equation}}}
   			which completes the proof of Proposition 1.
   			
   			\section{Proof of Proposition~2}
   			\label{proof_proposition_2}
   			We first prove that $\mathds{E}[{L}_J^{-}]$ increases with $x$.
   			Suppose $J \geq C_\mathrm{m} + 1$, and take the derivative of $\bar{h}$ with respect to $x$,
   			\begin{equation}
   				\begin{split}
   					\frac{\partial \mathds{E}[{L}_J^{-}] }{\partial x} & 
   					= L_\mathrm{m}\left[ \frac{\gamma(KC_\mathrm{m},x)}{(K C_\mathrm{m}-1)!} \frac{KC_\mathrm{m}}{x} \right]^2 + L_\mathrm{m} \left( \frac{x^{KC_\mathrm{m}-1}e^{-x}}{(KC_\mathrm{m}-1)!} \right)^2\\
   					& ~~+ L_\mathrm{m} \frac{\gamma(KC_\mathrm{m},x)}{(K C_\mathrm{m}-1)!}\frac{x^{KC_\mathrm{m}-1} e^{-x} }{(KC_\mathrm{m}-1)!} \left(1-\frac{KC_\mathrm{m}+1}{x}\right) \\
   					& = L_\mathrm{m} \left(\frac{\gamma(KC_\mathrm{m},x)}{(K C_\mathrm{m}-1)!} \frac{KC_\mathrm{m}}{x} -  \frac{x^{KC_\mathrm{m}-1}e^{-x}}{(KC_\mathrm{m}-1)!} \right)^2	\\
   					& ~~+ L_\mathrm{m} \frac{\gamma(KC_\mathrm{m},x)}{(K C_\mathrm{m}-1)!}	\frac{x^{KC_\mathrm{m}-1}e^{-x}}{(KC_\mathrm{m}-1)!} \left( 1+\frac{KC_\mathrm{m}-1}{x} \right).
   				\end{split}
   			\end{equation}
   			As $KC_\mathrm{m} \geq 1$, $\frac{\partial \mathds{E}[{L}_J^{-}] }{\partial x} > 0$.
   			In addition, $\frac{\partial \mathds{E}[{L}_J^{-}] }{\partial x} > 0$ can be proved in the same way for $J \leq C_\mathrm{m}$.
   			As $\mathds{E}[{L}_J^{-}]$ increases with $x$ for any given $J$, $\bar{h}$ increases with $x$ and Proposition~1 can be proved.
   			
   			\section{Proof of Proposition~3}
   			
   			\label{proof_proposition_3}
   			
   			{{According to (\ref{eq_waiting_time_18}), $\hat{W}_\mathrm{M} \leq \hat{T}_\mathrm{m}$ can be written as}}
   			{{\begin{equation}
   						\lambda'_\mathrm{v} \left(\frac{L_\mathrm{m}}{R_\mathrm{RM}}\right)^2 - 2  \left[1+ \lambda'_\mathrm{v} \hat{T}_\mathrm{m} \right] \frac{L_\mathrm{m}}{R_\mathrm{RM}} + 2 \hat{T}_\mathrm{m} \geq 0,
   					\end{equation}}}
   					which is equivalent to 
   					{{\begin{equation} 
   								\label{eq_R_RM_condition_1}
   								\frac{L_\mathrm{m}}{R_\mathrm{RM}} \leq \frac{1}{\lambda'_\mathrm{v}} \left(1+ \lambda'_\mathrm{v} \hat{T}_\mathrm{m} -\sqrt{\left(\lambda'_\mathrm{v} \hat{T}_\mathrm{m}\right)^2 +1} \right), 
   							\end{equation}
   							or 
   							\begin{equation}
   								\label{eq_R_RM_condition_2}
   								\frac{L_\mathrm{m}}{R_\mathrm{RM}} \geq \frac{1}{\lambda'_\mathrm{v}} \left(1+ \lambda'_\mathrm{v} \hat{T}_\mathrm{m} + \sqrt{\left(\lambda'_\mathrm{v} \hat{T}_\mathrm{m}\right)^2 +1} \right). 
   							\end{equation}}}
   							As $\lambda'_\mathrm{v} L_\mathrm{m} \leq R_\mathrm{RM}$ is required for the system stability, (\ref{eq_R_RM_condition_1}) needs to be guaranteed.
   							Thus, the RSU transmission rate requirement is given by Proposition~3.
   							
   							\section{Proof of Proposition~4}
   							\label{proof_proposition_4}
   							Denote by $\check{R}_\mathrm{RM}$ the minimal RSU rate requirement in Eq.~(\ref{eq_R_RM_result}) (i.e., the right side), and $z \triangleq \lambda'_\mathrm{v} \hat{T}_\mathrm{m} = (1-P_\mathrm{acc}) \lambda_\mathrm{v} \hat{T}_\mathrm{m}$.
   							Taking derivative of $\check{R}_\mathrm{RM}$, we have 
   							\begin{equation}
   								\begin{split}
   									\frac{\partial \check{R}_\mathrm{RM}}{\partial z} & = \frac{\sqrt{z^2+1}-1}{\left(1+z-\sqrt{1+z^2}\right)^2} \frac{1}{\sqrt{1+z^2}} \frac{L_\mathrm{m}}{\hat{T}_\mathrm{m}}, \\
   									& =  \frac{1-\frac{1}{\sqrt{z^2+1}}}{(1+z-\sqrt{z^2+1})^2} \frac{L_\mathrm{m}}{\hat{T}_\mathrm{m}} >0,
   								\end{split}
   							\end{equation}
   							since $z>0$.
   							In addition,
   							\begin{equation}
   								\begin{split}
   									& \frac{\partial^2 {\check{R}_\mathrm{RM}}}{\partial z^2} =   \frac{z(1+z^2)^{-\frac{3}{2}}(1+z-\sqrt{ 1+z-\sqrt{1+z^2}})}{(1+z-\sqrt{1+z^2})^3} \\
   									& ~~~~~~~\frac{{+ 2\left(\frac{1}{\sqrt{1+z^2}}-1\right)\left( 1-(1+z^2)^{-\frac{1}{2}} z\right)}}{}  \\
   									= & \frac{(z+1)\left[ -(2z^2-z+2) +2(z+1)\sqrt{1+z^2} \right]}{(1+z-\sqrt{1+z^2})^3 (1+z^2)^{\frac{3}{2}}}.
   								\end{split}
   							\end{equation}
   							As 
   							\begin{equation}
   								\begin{split}
   									& \left(2z^2-z+2\right)^2 -\left[ 2(z+1) \sqrt{1+z^2} \right]^2 \\
   									& = 4z^2+5 >0,
   								\end{split}
   							\end{equation}
   							and $z>1$,
   							$\frac{\partial^2 {\check{R}_\mathrm{RM}}}{\partial z^2} <0$.
   							Proposition 3 is proved.
   							
   							\section{Proof of Proposition~6}
   							\label{proof_proposition_6}
   							
   							We prove the content hit ratio decreases with $\rho$ when $\rho<1$.
   							For notation simplicity, define function $\zeta(\rho) = \frac{1-\rho^a}{1-\rho^A}$, where $\rho\neq 1$ and $A>a>0$.
   							Take derivative of $\zeta(\rho)$:
   							\begin{equation}
   								\begin{split}
   									& \frac{\partial}{\partial \rho} \zeta(\rho) 
   									= \frac{A \rho^{a-1}}{1-\rho^A} \left[\frac{1-\rho^a}{1-\rho^A}\rho^{A-a}-\frac{a}{A}\right].
   								\end{split}
   							\end{equation} 
   							Take derivative of $\frac{1-\rho^a}{1-\rho^A}\rho^{A-a}$:
   							\begin{equation}	
   								\label{eq_deri_1}
   								\frac{\partial}{\partial \rho} \left(\frac{1-\rho^a}{1-\rho^A}\rho^{A-a}\right) = \frac{(A-a)-A \rho^a +a\rho^A}{(1-\rho^A)^2} \rho^{A-a-1} > 0,
   							\end{equation}
   							as $\frac{\partial}{\partial \rho} \left[(A-a)-A \rho^a +a\rho^A\right] = Aa \rho^{a-1} \left(\rho^{A-a}-1\right) \leq 0$, and $\lim\limits_{\rho \rightarrow 1} (A-a)-A \rho^a +a\rho^A = 0$.
   							Therefore, 
   							\begin{equation}
   								\begin{split}
   									& \min \limits_{\rho} \frac{1-\rho^a}{1-\rho^A}\rho^{A-a} = \lim\limits_{\rho \rightarrow 0} \frac{1-\rho^a}{1-\rho^A}\rho^{A-a}  = 0,
   								\end{split}
   							\end{equation}
   							and
   							\begin{equation}
   								\begin{split}
   									& \max\limits_{\rho} \frac{1-\rho^a}{1-\rho^A}\rho^{A-a} = \lim\limits_{\rho \rightarrow 1} \frac{1-\rho^a}{1-\rho^A}\rho^{A-a}  \\
   									& = \lim\limits_{\rho'\rightarrow 0} \frac{1-(1-\rho')^{\frac{a}{A}}}{\rho'}  = \lim\limits_{\rho'\rightarrow 0} \frac{\mbox{d}\left(1-(1-\rho')^\frac{a}{A}\right)}{\mbox{d}\left(\rho'\right)} \\
   									& = \lim\limits_{\rho'\rightarrow 0} \frac{a}{A} \left(1-\rho'\right)^{\frac{a}{A}-1} = \frac{a}{A},
   								\end{split}
   							\end{equation}
   							where $\rho' = 1-\rho^A$.
   							Therefore, $\frac{\partial }{\partial \rho} \zeta(\rho)$ is negative for $0<\rho<1$.
   							Similarly, we can prove $\frac{\partial }{\partial \rho} \zeta(\rho) < 0$ for $\rho>1$, and the details are omitted due to page limit.

\bibliographystyle{IEEEtran}

\end{document}